# Interdigitated Columnar Representation of Personal Space and Visual Space in Human Parietal Cortex

Roger B. H. Tootell [2,3,4] *, Zahra Nasiriavanaki [1,2]*, Baktash Babadi [1,2], Douglas N. Greve [2,3,4], Shahin Nasr [2,3,4], and Daphne J. Holt [1,2,3]

1. Department of Psychiatry, Massachusetts General Hospital, Charlestown, MA, USA, 02129
2. Harvard Medical School, Boston, MA, USA, 02115
3. Athinoula A. Martinos Center for Biomedical Imaging, Massachusetts General Brigham Hospital, Boston, MA, USA, 02129
4. Department of Radiology, Massachusetts General Hospital, Charlestown, MA, USA, 02129

* Equal contribution

Corresponding author:   Roger B. H. Tootell

Department of Radiology, Massachusetts General Hospital,

149 13th St., Charlestown, MA, USA, 02129

tootell@mgh.harvard.edu

Number of pages: 52

Number of Figures: 10

Number of tables: 1

Abstract: 245 words

Introduction: 646 words

Discussion: 1417 words

## Conflict of Interest

The authors declare no competing financial interests.

## Acknowledgements

This research was supported by grant NIMH RO1MH109562 (DJH). We thank Dr. Jonathan Polimeni for help with the MR sequences.

## Abstract

Personal space (PS) is the distance that people prefer to maintain between themselves and unfamiliar others. Intrusion into a given person's PS typically evokes discomfort and an urge to move away. Physiological evidence suggests that responses to threatening intruding stimuli involve parietal cortex. Based on this and related evidence, we hypothesized that the encoding of interpersonal distance is transformed from purely sensory to PS-centered within some sites in parietal cortex. This hypothesis was tested using 7T fMRI at high spatial resolution. In response to different visual stimuli across a range of virtual distances, we found two categories of distance encoding in two corresponding columns within parietal cortex. One set of columns ("P columns") responded selectively to moving and stationary face images presented at virtual distances that were nearer (but not further) than each subjects' behaviorally-defined PS boundary. In the majority of P columns, BOLD response amplitudes increased monotonically and nonlinearly with increasing virtual face proximity. In a complementary subset of P columns, BOLD responses decreased with increasing proximity. A second set of parietal columns ("D colummns") responded selectively to disparity-based 'near' or 'far' distances, in random dot stimuli, similar to disparity-selective columns described previously in occipital cortex. Importantly, in parietal cortex, maps of P columns were systematically non-overlapping (interdigitated) with D columns. These results suggest that the transformation of spatial information, from visual- to body- (or person-) centered, may be computed in multiple small sites, rather than across a larger cortical gradient, in parietal cortex.

**Significance Statement**

Recent COVID-related social distancing practices have emphasized the need to better understand brain mechanisms which regulate "personal space", i.e. the closest interpersonal distance that is comfortable for an individual. Using high spatial resolution brain imaging, we tested whether a map of external space is transformed from purely visual (3D-based) information to a map of the whole body ('person'), in parietal cortex. This transformation was mediated by two sets of columns: one set encoded interpersonal distance and the other encoded visual distance, binocularly. These two sets of columns are mutually segregated in parietal cortex. These results suggest that the cortical transformation of sensory-to-social encoding of space near the body involves multiple local processes within parietal cortex.

## Introduction

The importance of personal space regulation has been highlighted recently by the widespread adoption of 'social distancing' practices, designed to reduce COVID-19 transmission (Cartaud et al., 2020, Iachini et al., 2020, Welsch et al., 2020, Holt et al., 2021, Welsch et al., 2021). However, well before the introduction of this deliberate social distancing, many behavioral studies described the fundamental discomfort that people experience when an unfamiliar person becomes 'too close', intruding into their personal space (PS) (Hayduk, 1978, Hayduk, 1983). This interpersonal distancing between conspecifics may be related to the survival-based distancing that occurs between prey and predator species (Hediger, 1955, Ardrey, 1966).

Based on distance from another person, most models of PS processing posit an inner zone that is close to the body and a larger surrounding zone, e.g. the distinction between 'intimate/personal' and 'social/public' zones (Hall, 1966), with an intervening personal space boundary. Repeated measurements of PS boundary distance are highy reliable *within* a given individual, but vary significantly *between* individuals (Hayduk, 1983, Bar-Haim et al., 2002, Cléry and Hamed, 2018, Tootell et al., 2021, Welsch et al., 2021). Within the inner (PS) zone, discomfort increases gradually with closer interpersonal proximity (Hayduk, 1981a, Welsch et al., 2019, Tootell et al., 2021). Beyond the PS, discomfort is typically not elicited by the presence or absence of another person (but see (Welsch et al., 2019)). Analogously, skin conductance (a measure of 'arousal') increases when experimental confederates are positioned closer to a subject, within (but not beyond) the PS (Tootell et al., 2021, Candini et al., 2021). The concept of personal space (emphasizing the discomfort and/or threat response evoked by interpersonal intrusion) is closely related to the concepts of peripersonal space (PPS, emphasizing the body limits) (Bogdanova et al., 2021), body ownership (BO), body schema, and other distinctions.

In contrast, much less is known about the brain mechanisms that underlie PS processing. Several human fMRI studies have suggested a role for parietal cortex in processing of PS and/or BO (Lloyd et al., 2006, Schienle et al., 2015, Grivaz et al., 2017). Another study (Holt et al., 2014) described relatively higher fMRI activity in response to images of approaching (compared to withdrawing) faces, in two specific parietal regions. Because these responses were stronger to approaching faces than withdrawing objects, these results were interpreted as a possible response to personal space intrusion, rather than to looming stimuli *per se*.

Some involvement of parietal cortex in PS processing is not unexpected because: 1) PS intrusion is a specific type of threatening event, and 2) in monkeys, threat-defensive responses have been reported in parietal cortex (and in premotor areas F4/5, which are interconnected with parietal cortex) (Luppino et al., 1999). For instance, defensive behavior was repeatedly elicited by electrical micro-stimulation in inferior parietal cortex (Cooke et al., 2003), and (at lower threshold) in areas F4/5 (Graziano et al., 2002).

The processing of threats in the visuo-spatial environment might benefit from a transformation of distance encoding which enhances the representation of the threatening stimuli. Specifically, here we tested whether/how the *eye-centered* spatial representation (from occipital cortex) is transformed into a *body-centered* (more egocentric) representation in inferior parietal cortex, as suggested earlier (Hadjidimitrakis et al., 2014, Cléry et al., 2017).

Such a hypothetical transformation between eye- to body-centered spatial representations could be topographically organized in different ways, including: 1) a large-scale gradient distributed across the parietal surface, 2) in multiple smaller-scale local networks (e.g. cortical columns), or 3) without systematic topographic interrelationships. Such hypothetical alternative organizations imply correspondingly differing computational schemes. Our results generally support the second hypothesis. Specifically, the current results demonstrated multiple, local transformations of spatial information in parietal cortex, mediated via two sets of cortical columns, which differ in the nature of their spatial- (and threat-) encoding properties, ranging from visual to more egocentric scales. To our knowledge, no type of cortical column has been reported previously in parietal cortex.

**Experimental Design and Statistical Analysis**

Participants

Seven human subjects (four females; age range: 22–29 years; mean: 24.57) participated in this study. All subjects had normal or corrected-to-normal visual acuity (Snellen test) and radiologically normal brains, without any history of neuropsychiatric disorders. Written informed consent was obtained from all subjects before enrollment, in accordance with the Declaration of Helsinki. All experimental procedures were conducted according to Massachusetts General Hospital protocols, and approved by the Massachusetts General Brigham Institutional Review Board.

Measurements of personal space

In each subject, the size of personal space (i.e., preferred interpersonal distance) was measured in two ways, prior to scanning. First, we used the standard 'Stop Distance Procedure' (SDP) (Williams, 1971, Hayduk, 1978, Hayduk, 1983), in which each subject identified their preferred, comfortable distance from an 'intruder' (an experimental confederate), in a neutral laboratory setting. During this procedure, subjects were asked to stand still, facing the intruder, who initially stood 3 m away from the subject. Subjects were instructed as follows: "*Please stand still as my colleague walks slowly towards you. Say 'okay' when they reach a distance that you would normally stand to talk to someone who you have just met, when you start to feel slightly uncomfortable. When you say 'okay', they will pause. Please make sure to maintain eye contact with my colleague throughout the procedure*". The distance between the subject and the intruder was measured after each trial, defined as the distance between the proximal tips of their shoes on the floor (the 'stop distance').

The second set of measurements was acquired while each subject lay prone in the 7T scanner, without any scanning. Each subject performed a computerized version of the SDP, in which subjects were presented with face images which appeared to move towards them. A small central fixation target was superimposed on the bridge of the nose of each face image. The subjects were instructed to maintain fixation, and to stop the changes in virtual distance in the face image (by pressing the button on the button box) when the face reached a virtual distance at which the subject began to feel uncomfortable, i.e. at a distance where they *"would have a conversation with someone they have met for the first time"*.

The outside-scanner SDP values were generated as a standard for comparison to values in the literature, and to the within-scanner measurements. In addition to these purposes, the within-scanner values were also used to pre-calculate stimulus dimensions, and for the data analysis. In analyses of both types of personal space measurement, the first trial was considered a 'warm up' trial, as described previously (Hayduk, 1981b), and excluded from the main analyses.

Our use of face images as stimuli here is consistent with previous studies which have also presented human-like stimuli, including mannequins, VR-based avatars, and faces images (Bailenson et al., 2003, Mccall et al., 2009, Llobera et al., 2010, Rinck et al., 2010, Wieser et al., 2010, Iachini et al., 2014, Schienle et al., 2015, Iachini et al., 2016, Ruggiero et al., 2016, Hecht et al., 2019, Welsch et al., 2019, Tootell et al., 2021) to elicit responses analogous to those evoked by real human subjects.

Visual stimuli

Stimuli were presented via a LCD projector (1024 x 768-pixel resolution, 60 Hz refresh rate) onto a rear-projection screen, viewed through a mirror mounted on the receive coil array. Matlab 2018b (MathWorks) and the Psychophysics Toolbox were used for stimulus presentation (Brainard, 1997, Pelli, 1997).

Experiment 1: Approaching vs. withdrawing 3D face stimuli

A series of sixteen face images were presented, in four different conditions: Female Face Approach, Female Face Withdrawal, Male Face Approach and Male Face Withdrawal. Face stimuli of neutral emotional valence were generated using a commercially available program (FaceGen: http://www.facegen.com). Each approaching or withdrawing face series was presented throughout a 15-second long block, in a counterbalanced order, with eighteen blocks/run, and eight runs/session. To define baseline BOLD levels, we also presented two 15-second blocks/run of uniform grey fixation blocks, one at the beginning and one at the end of each run.

The stimulus face images changed in size and binocular disparity, within a wide field of view, matching the corresponding visual changes that would occur during the perception of a real person, either approaching or withdrawing, across a range of 39-300 cm from the viewer, at a constant walking speed (here, 1.12 m/sec). Stimuli were similar to those used previously (Holt et al., 2014), except that 3-D cues were added to the face images, based on differences in binocular disparity. Subjects viewed all disparity-varying stimuli using red-cyan anaglyph glasses (Kodak Wratten filters No. 25 (red) and 44A (cyan)) that were mounted on the top of the receive head coil, inside the scanner.

Subjects performed a dummy attention task during the presentation of the face stimuli, in which a blue dot was presented repeatedly within each block each time at a random location of the computer screen (duration = 400 ms; three dot presentattions in each 15-sec block, timed unpredictably relative to the main stimuli). Subjects were instructed to maintain fixation on a small target located at the center of the images, and to press a button (located on the button box inside the scanner) whenever they saw a blue dot presented on the screen. The instructions were repeated before each run.

Experiment 2: Stationary face stimuli

In Experiment 2, subjects viewed stationary face images, using a subset of the images that were presented during Experiment 1. Here, each stationary face was presented at different virtual distances, in a pseudorandom order, in an event-related design. Each face was presented for 3 seconds followed by an inter-trial interval (a spatially uniform gray screen) varying between 3 – 12 seconds. Virtual face distances were individually pre-calculated for each subject, based on each subject's personal space values acquired previously within the scanner (see above). Each face was presented at nine systematically varied virtual distances: four distances inside the subject's personal space boundary, four distances outside the subject's personal space boundary, and one distance corresponding to the subjects' personal boundary; at 37, 57, 76, 91, 100, 182, 295, 419, and 524% of each subjects' personal space boundary.

Experiment 3: Binocular disparity (stereoscopic) stimuli

In this experiment, disparity-varying stimuli were presented in sparse (5%) random dot stereograms based on small red or green squares (each 0.09° x 0.09°) presented against a black background, which extended 20° x 20° in the visual field. Two random dot stimuli (each either red or green) were overlaid on the display screen, and binocularly fused during BOLD acquisitions. In the main condition, stimuli formed a stereoscopic percept of a regular array of cuboids that varied sinusoidally in depth, i.e. within a 0.2-0.4 degrees of horizontal binocular disparity, stereoscopically 'near' or 'far', relative to the fixation plane, with independent phase, similar to stimuli described earlier (Tsao et al., 2003, Nasr et al., 2016). In the control condition, the fused percept spanned a fronto-parallel plane intersecting the fixation target (i.e., zero depth at the point of ocular convergence). Each block-designed experimental run included eight stimulus blocks (24 s/block), beginning and ending with an additional 12 s/ of uniform gray ("blank"). When stereoscopically fused, this random dot stimulus produced a percept of multiple 3D cuboids, each of which moved continuously towards then away (and vice versa) from the subject. When viewed monocularly, or when the binocular images were identical, these control stimuli instead appeared as an array of dots, moving coherently and slowly from left to right (and vice versa) in the zero disparity plane, uncorrected for variation in the arc tangent.

To measure and control the level of attention during the 3D disparity scans, subjects were required to report changes in the shape (identifying the longer axis) of a small (average length/width = 0.1- 0.2º) rectangular fixation target by pressing a button located on the button box inside the scanner.

Imaging

All experiments were conducted in a 7T Siemens whole-body scanner equipped with SC72 body gradients (maximum gradient strength, 70 mT/m; maximum slew rate = 200 T/m/s) using a 32-channel helmet receive coil array and a birdcage volume transmit coil. The anatomical scan collected in each subject was based on a high-resolution multi-echo T1-weighted magnetization-prepared gradient-echo image (MPRAGE, 0.80 mm isotropic voxels, with a repetition time (TR) of 2530 ms, an inversion time of 1100 ms, an echo time (TE) of 1.76 ms, a flip angle (FA) of 7º and a field of view (FOV) of 240 mm). Functional images were acquired using a single-shot gradient-echo EPI with 1.1 mm (isotrpic) voxel size, a TR of 3000 ms, a TE of 26 ms, a flip angle of 90°, with 90 frames in Experiment 1 and 2, and 80 frames in Experiment 3, spanning a region including dorsal and posterior occipital cortex, and the dorsal extent of parietal cortex.

In each subject, the number of functional volumes acquired was 1440 for Experiment 1, 720 for Experiment 2, and 1280 for Experiment 3.

General imaging data analysis

All structural and functional MRI data analyses were performed using the Freesurfer Functional Analysis Stream (FSFAST) v.6 (https://surfer.nmr.mgh.harvard.edu/fswiki/FsFast). In four main steps: 1) bias field correction, 2) cortical reconstruction, 3) preprocessing of the functional data, and 4) first level and group statistical analyses (Dale and Sereno, 1993, Dale et al., 1999, Fischl and Dale, 2000, Fischl et al., 2001, Fischl et al., 2002, Fischl et al., 2004a, Fischl et al., 2004b).

*Bias field correction:* To correct the anatomical scan for signal intensity variation across the brain volume caused by inhomogeneity of the B1 field at 7T, a bias field correction was performed using SPM 8. A Bayesian model estimated a smooth function that was multiplied with the image using prior knowledge about the field distributions likely to be encountered (Ashburner and Friston, 2005). The model assumed that the inhomogeneity of the field (a type of 'noise') is multiplicative and due to variations of the tissue properties in each voxel, rather than due to noise from the scanner. The full-width-half-maximum (FWHM) for the Gaussian smoothness used a bias of 18 mm, with a very light (0.0001) bias regularization (Zaretskaya et al., 2018).

*Cortical reconstruction:* The automated reconstruction of the structural data (Freesurfer (https://surfer.nmr.mgh.harvard.edu)) included the following steps: image realignment and motion correction, skull stripping (removing the non-brain tissue), gray-white matter segmentation, reconstruction of cortical surface models, and labeling of the regions on the cortical surfaces based on probabilistic information derived from a manually labeled training set.

*fMRI Processing:* BOLD data were registered to the same-subject anatomical data using Boundary Based Registration (Greve and Fischl, 2009) and corrected for head motion, then sampled onto the cortical surface, where all further analysis was performed. Data were not spatially smoothed (i.e., 0 mm FWHM). The hemodynamic response amplitudes for each condition were estimated using a general linear model in FSFAST; regressors were constructed by convolving the stimulus box car with the SPM canonical hemodynamic response function. Head motion parameters were treated as nuisance regressors; polynomial regressors were included to account for low frequency drift. Voxel-wise statistical tests were conducted by computing contrasts.

*Quality control:* All brain images were visually inspected for brain coverage, registration quality and anatomical defects. The FSL motion outlier algorithm was used to eliminate time points showing excessive head motion across volumes (Power et al., 2012), by using a compound matrix of outlier timepoints as a regressor in the GLM. Among the several metrics available for defining the outlier time point (Power et al., 2012, Bastiani et al., 2019), we used the DVARS metric (D referring to temporal derivative of time courses, VARS referring to RMS variance over voxels) (Smyser et al., 2010), which reflects the rate of change of the BOLD signal across the entire brain in comparison to the previous time point (Power et al., 2012). The threshold used to define an outlier was set to the default value suggested by Power et al (Power et al., 2012) (75th percentile + 1.5 times the Inter Quartile Range of the DVARS metric). Any timepoint with a value larger than this threshold was eliminated from further analysis, as an outlier. Very few timepoints (0.08% of the total) were rejected based on this criterion.

In addition to time point motion correction, the six rigid-body motion parameters within each volume were used as regressors, to account for absolute motion. Based on the threshold criterion above, 9.9% of the timepoints were excluded, across all the runs (n=264) from all three experiments.

In the responses to the dummy attention task, outliers were defined as a score (in any of the experiments) of less than two standard deviations from the mean. However, no subjects' data met this criterion (Experiment 1: M = 92.15%, SD = 3.3; Experiment 3: M=93.48%, SD=7.09).

*Region-of-Interest (ROI):* In prior group-averaged, conventional-resolution fMRI studies, preferential activity to both approaching (Holt et al., 2014) and disparity-selective (Tsao et al., 2003) stimuli were found within the dorsal medial and lateral parietal sulci. Accordingly, the ROI here was based on anatomical (sulcal/gyral) features bounding this region, specifically including the dorsal half of the dorsal posterior occipital sulcus (DPOS) on the medial bank, continuing to the lateral bank on the inferior parietal sulcus (IPS), then crossing perpendicularly to the small sulci on the lateral bank of the IPS, to intersect the starting point near DPOS. In the single subject data, this ROI was defined on the cortical surface of each subject. In the group averaged data, this ROI was defined based on the averaged gyri and sulci in the group-averaged cortical surface (Fischl et al., 1999).

Specific analysis of 7T data

*Consistency across sessions.* To quantify the level of consistency between approach- and withdrawal-biased activity across different scan sessions, we measured the fMRI signal change evoked by the contrasts of interest, at a threshold of $p < 0.05$ (Face Approach > Face Withdrawal and Face Withdrawal > Face Approach), across the first two scan sessions of each experiment, for each vertex within our ROI, and also in the average of the two scan sessions. The level of overlap in the resultant maps was used to quantify the consistency across sessions, when aligned (100% overlap = perfect consistency), compared to systematically non-aligned. These control measurements of non-aligned maps were calculated by pseudo-randomizing the alignment of maps from session 1 relative to session 2.

*Consistency across depth.* To assess the radial ('columnar') consistency in BOLD-activated sites, we subdivided and compared fMRI activity centered at multiple cortical depths, as described previously (Polimeni et al., 2010). For each subject, one surface was generated at the gray matter–white matter interface (depth = 1.0), and a second surface was generated at the superficial boundary between the gray matter and the pia (depth = 0.0), based on each subjects' high-resolution structural scans (see above) using FreeSurfer (Dale et al., 1999, Fischl et al., 2002). In addition to these two bounding surfaces, intermediate "mid-gray" surfaces were also generated at 50% of the depth of the local gray matter (depth = 0.5, including cortical layer 4). A "lower layer" surface was also generated at 80% of the depth of the local gray matter (depth = 0.8, including layers 5 and 6) (Dale et al., 1999).

*Activity maps from Experiment 1 (Approach/withdrawal).* To create group-averaged maps, cortical surfaces from all hemispheres and subjects (14 hemispheres) were registered to an average reference surface for each hemisphere, and functional data were registered onto the average anatomical data, corrected for head motion and normalized without spatial smoothing (FWHM=0) (Fischl et al., 1999). Statistical analyses were performed by fitting a univariate general linear model (GLM) with a one-sample group mean design (OSGM), using a weighted least square to the fMRI data. The significance maps were created based on the predefined conditions, at a threshold of $p < 0.01$.

*Columnar organization.* To examine the BOLD maps from a viewpoint perependicular to the cortical topography, the fMRI signal change evoked by the functional contrast of interest (face approach vs. face withdrawal) was sampled at multiple equally-spaced points across the cortical depth. That activity was then illustrated along a line drawn along each subject's cortical surface map, which crossed patches of high and low activity.

Next, a more formal analysis was conducted to test for a columnar organization across all our subjects, and within all our parietal ROIs, by calculating the correlation between the selective activity in vertices distributed along each of the three cortical axes in the flattened cortical maps: X and Y within the cortical map, and Z along the perpendicular (radial) axis. As a control, we measured BOLD activity at two different cortical depths: superficial (depth = 0.1) and deep (depth = 0.9). Given the average cortical thickness within our ROI (2.37 mm), corresponding points on these two surface maps were centered 1.9 mm apart in depth, on average. Then, for each of the three axes, correlation values were calculated independently for each subject, and compared using t-tests following transformation of the Pearson r values to z values using Fisher's transformation formula.

*BOLD responses from Experiment 2 (Stationary Faces):* For each subject, the average BOLD responses within the approach- and withdrawal-selective columns were calculated for each virtual stimulus distance. These distances were calculated separately for each subject, as percentages of the same subject's mean personal space boundary, measured in the scanner (see above).

*Overlap Measurements of the approach/withdrawal vs. stereopsis activity maps.* In several analyses, we measured the extent of overlap in evoked activity in different functional contrasts in our parietal ROI. For instance, we measured the overlap in the maps from each of four functional contrasts (approach-biased, withdrawal-biased, stereo near-biased and stereo far-biased), independently in the approach/withdrawal and disparity-selective columns, across a wide range of thresholds (log (p) 1.3 -10). The overlap was calculated as a percentage of the surface area, in two ways: 1) when the maps were correctly aligned, and 2) (as a control condition), when one of the maps was randomly misaligned relative to the other (i.e., spatially shuffled) 1000 times.

**Codes/Software**

Behavioral and brain data (available upon request).

MATlab (RRID: SCR_001622; https://www.mathworks.com).

Freesurfer (RRID:SCR_001847; https://surfer.nmr.mgh.harvard.edu/fswiki/FsFast).

SPM (RRID:SCR_007037; https://www.fil.ion.ucl.ac.uk/spm/).

SPSS (RRID:SCR_007037; https://www.fil.ion.ucl.ac.uk/spm/).

Psychophysics Toolbox (RRID:SCR_002881; http://psychtoolbox.org/docs/ Psychtoolbox).

**Results**

General procedures

First, we measured the size of personal space (preferred interpersonal distance) in each subject (n = 7). Then each subject was scanned multiple times, on different days, in a high field MRI scanner (7T whole-body system, Siemens Healthcare) to localize fMRI activity evoked by approaching-vs.-withdrawing face images (two scan sessions) and random dot arrays varying in binocular disparity (stereopsis) (two sessions). Five of the seven subjects (three females) participated in an additional experiment, which tested responses to stationary faces that were presented at a range of distances (one session each).

Behavioral measurements of personal space size

Figure 1 shows the personal space size measured for each subject using: A) the standard Stop Distance Procedure (SDP) (Williams, 1971, Hayduk, 1981a, Hayduk, 1983), in response to intrusion by a human confederate, outside of the scanner, and B) in a modified, within-scanner SDP, in response to the 3D face images (Figure 1B) (see Methods). A repeated measure ANOVA showed no differences between the within-subject SDP measurements (Greenhouse-Geisser $F_{(2.5, 16.1)} = 0.95$, $p = 0.45$, mean = 57.31 cm, SD = 11.3). Similarly, in the within-scanner personal space data, a repeated measure ANOVA showed statistically equivalent mean values over all repeated measurements. We found no significant habituation across trials (Greenhouse-Geisser $F_{(2.9, 17.8)} = 1.85$, $p = 0.17$; mean = 60.72 cm, SD = 15.01).

These behavioral results are consistent with results from prior studies which reported high within-subject reliability over time in the standard SDP (Hayduk, 1981a, Hayduk, 1983, Wormith, 1984, Welsch et al., 2021, Tootell et al., 2021). This demonstration of high reliability in personal space measurements was an important prerequisite for the analyses described below, because our statistical sensitivity relied on extensive signal averaging over repeated measurements.

Within-scanner task response accuracy

Mean accuracy on the dummy attention task (all subjects, all sessions) was 92.2% in response to the face stimuli, and 93.5% in response to the random dot stimuli.

Experiment 1: Approach-selective activity

In a prior study based on group-averaged (n=22) conventional fMRI results (3T, 3 mm isotropic), presentation of approaching versus withdrawing face images consistently produced two patches of approach-selective activity in dorsal parietal cortex, one located posterior to the other (Holt et al., 2014). Using similar but more realistic face stimuli here (see Methods), we tested for comparable maps of activity using a 7T scanner, in both group-averaged (n=7) and in single subject analyses, both at high (1.1 mm isotropic) spatial resolution.

Consistent with the earlier 3T results, the activity evoked here included two regions of approach-dominated activity, in anterior and posterior sites in the parietal cortex (Figure 2). Our region of interest (ROI) was located in the inderior parietal cortex (PPC), immediately anterior to (and known to share connections with) dorsal occipital ('visual') cortex. Based on its proximity to visual cortex, we hypothesized that this inferior parietal ROI would be most likely to show evidence for a transformation from eye-based spatial encoding (from visual cortex) to body-based spatial encoding (in parietal cortex).

A prior 3T study (Quinlan and Culham, 2007) found that a discrete site within the dorsal occipital sulcus (DPOS) responded selectively to 'near' (compared to 'far') visual objects. Here, the corresponding parietal site (white arrows in Figure 2B) also showed a strong bias for the approaching (compared to withdrawing) stimuli, in the averaged activity, and in 11 (of 14) hemispheres tested.

Single subject maps

Overall, the activity maps appeared more 'patchy' in individual hemispheres (Figure 3A,B), compared with the group-averaged map of the same data (Figure 2). It could be argued that this 'patchiness' was due to statistical noise in the individual maps. However, observations suggest that the 7T maps were highly consistent within each subject/hemisphere, when compared across sessions in an overlap analysis (Figure 4A-F). Consistent with the initial observations, we found that the maps showed significantly more overlap across sessions within a given subject when the maps of the two sessions were topographically aligned, compared to the non-aligned (systematically shuffled) maps (t-test: p<0.0001) (Figure 4G-H). Thus the approach-versus-withdrawal maps did not appear to be dominated by statistical noise across sessions.

Within the parietal ROI, the specific topography of the evoked patches varied between subjects, akin to the idiosyncratic (fingerprint-like) mapping of cortical columns in macaque visual cortex, which also varies widely between individuals (Tootell et al., 1988, Horton and Hocking, 1996). Presumably, the idiosyncractic patches in each individual subject were effectively 'averaged out' in the group averaged maps here (Figure 2) and previously (Holt et al., 2014), because the patches in different subjects do not align with each other in the group-averaged maps.

In addition to patches that showed *approach*-biased activity, the single subject maps revealed patches that were more strongly activated by the *withdrawing* stimuli (Figure 3). Like the approach-biased patches, those withdrawal-biased patches remained largely consistent across scan sessions (Figure 4). However, the number of these withdrawal-biased patches was significantly lower than the number of the approach-biased patches, above all thresholds between $10^{-2}$ through $10^{-5}$ (all p < 0.01). In some (but not all) subjects/hemispheres, the topographical distribution of the withdrawal-biased patches favored the posterior extent of parietal cortex, in our small (n = 7) group average.

BOLD evidence for a columnar organization

These results raise the possibility that these stimuli differentially activated a previously unknown set of functionally distinct cortical columns in posterior parietal cortex. If the evoked 'patches' are indeed 'columns', and to the extent that BOLD activity faithfully reflects the 3D shape of neural activity, then these BOLD patches should be elongated along the radial axis (i. e. perpendicular to the cortical surface) within the gray matter. This radial consistency of a given functional property is a defining property of cortical columns (Mountcastle, 1957, Hubel and Wiesel, 1962, Hubel and Wiesel, 1968, Horton and Adams, 2005).

To test for such a radial elongation, we first examined the single subject data qualitatively, then conducted a formal statistical analysis.

First, Figure 5 suggests that the flattened activity maps that were sampled preferentially from the upper cortical layers were strikingly similar to those sampled from lower cortical layers, as in the well studied columns within visual area V2 (Nasr et al., 2016). In 3D, this supports a radial elongation in the BOLD-evoked patches.

Next we sampled the evoked activity differences from a perpendicular viewpoint, as in classic views of 'columns' across the cortical layers (see Methods). These results (Figure 6) further supported a radial elongation of the evoked BOLD responses, in response to both approaching and withdrawing stimuli.

Finally, we measured the correlation levels of stimulus-driven BOLD differences in the *surface-normal* (i. e. radial) versus *surface-parallel* (i. e. within-laminar) axes (see Methods). As a further control, two surface-parallel maps were tested: one at shallow cortical depths, and one at deeper depths (Figure 7). Consistent with the above observations (Figures 5,6), all fourteen hemispheres showed a significantly higher correlation in the approach-vs.-withdrawal functional contrast when sampled along the radial axis, compared to functional variation along either of the surface-parallel axes (Figure 7; all $p < 0.001$).

In fact, the correlation showed an even stronger radial bias in parietal cortex here, compared to that in the well-established columns in visual area V2 (cf. V2 in Figure 7 of (Nasr et al., 2016) with Figure 7 here). Specifically, the average z-scores in the current parietal cortex data measured in radial vs. surface-parallel axes were 0.84 vs. 0.26 respectively, whereas comparable values in the prior V2 data were 0.81 vs. 0.51. Given this consistent radially-elongated cortical organization of a common response function in the current results, below we refer to both the approach- and withdrawal-selective 'patches' as 'columns', specifically termed 'P' columns (i.e. presumptively related to personal space).

Experiment 2: Responses to stationary stimuli across a range of interpersonal distances

In the SDP (in both the classic version (Williams, 1971, Hayduk, 1983), and as adapted here for use in the scanner), the stimuli moved continuously, either approaching or withdrawing. Evidence from human (Cléry et al., 2015, Candini et al., 2021) and non-human primates (Cléry et al., 2017) raise the possibility that these dynamic personal space measurements reflect contributions from visual motion cues in addition to personal space *per se*, e.g. based on anticipation of the oncoming motion trajectory. However, it is also known that the discomfort evoked by personal space intrusion at a given distance is largely time-invariant, i.e. relatively constant over a wide range of inter-test intervals (seconds through months) and extensive repetitions (Hayduk, 1981a, Wormith, 1984, Tootell et al., 2021, Welsch et al., 2021) (see also Figure 1).

To eliminate such possible contributions of stimulus motion to measurements of personal space-related neural responses, Experiment 2 measured fMRI responses to *stationary* face stimuli. The stationary faces (a small subset of the face images used in Experiment 1) were presented at several virtual distances from the observer (Figure 8A), spanning both near and far distances relative to each subjects' personal space boundary (9 distances total), and at each subjects' personal space boundary, in pseudo-randomized order in an event-related fMRI design (see Methods). BOLD responses to the stationary stimuli were measured independently in the approach- and withdrawal-selective columns.

The dominant approach-selective columns showed a consistent pattern of BOLD responses to variations in the virtual distance (Figure 8B). The mean BOLD response to faces presented at the four largest distances (i.e., 'further' than the personal space boundary) showed a negligle (baseline) response (all p-values > 0.16), essentially non-responsive to the virtually 'far' face stimuli. In contrast, BOLD responses to faces presented closer than the personal space boundary increased steeply and monotonically, such that the highest response was evoked by the closest face. The withdrawal-selective columns showed a less dramatic response function to the stationary faces, which was nevertheless largely inverted in sign relative to baseline, and negligible at distances further than the personal space boundary (Figure 8B).

Modeling of approach vs. withdrawal columns

It has been reported that intrusion into personal space evokes higher fMRI levels in the amygdala (Kennedy et al., 2009, Todd and Anderson, 2009, Adolphs, 2010, Wabnegger et al., 2016); (see also (Nacewicz et al., 2006)), consistent with the long-standing hypothesis that the amygdala is involved in the processing of social threat and avoidance (Klüver and Bucy, 1937, Whalen et al., 1998, Emery et al., 2001, Mason et al., 2006). Thus here, it is possible that input from the amygdala limits the parietal cortical responses to stimulus distances that are closer (but not further) than each subject's personal space boundary (e.g. Figure 8B).

Alternatively, we hypothesized that a differential weighting of purely visual inputs could instead account for these response differences in parietal cortex. To evaluate the computational feasibility of this idea, we tested a scheme in which both approach- and withdrawal-selective population responses resulted from ascending ('upstream') input from visual cortex. The model matched the experimental results well (c.f. Figure 8B, C). Thus, the different response profile of the approach- and withdrawal-biased columns could reflect correspondingly different strengths of the local excitatory and inhibitory synaptic strengths from visual cortex to each column type in parietal cortex, without requiring additional input from the amygdala. Further testing is necessary to validate this model.

Experiment 3: Near- and far-selective columns based on binocular disparity

As in real life conditions, and as in the face images used in Experiments 1 and 2, the binocular disparities across the face stimuli varied with distance from the viewer. Thus it might be argued that the approach- and withdrawal-selective columns described above might involve disparity-selective neurons and columns, like those described previously in occipital visual cortex in macaques (Felleman and Van Essen, 1987, Poggio et al., 1988, Roy and Narahashi, 1992, Deangelis et al., 1998, Deangelis and Newsome, 1999, Adams and Zeki, 2001, Tanabe et al., 2005, Chen et al., 2008) and in humans (Goncalves et al., 2015, Nasr et al., 2016, Tootell and Nasr, 2017). However, this argument presumes that 'near' and 'far' disparity-selective columns extend beyond occipital cortex, anteriorly into parietal cortex – and no such hypothetical disparity columns have been reported previously in parietal cortex.

Accordingly, we next tested for the presence of disparity columns in the posterior parietal ROI, using random dot stereogram stimuli that selectively activated 'near' and 'far' distance disparity-selective columns in past studies of occipital cortex (Tsao et al., 2003, Nasr et al., 2016, Nasr and Tootell, 2018, Nasr and Tootell, 2020). When disparities in the two monocular images were binocularly fused, this random dot stimulus produced a stereoscopic percept of multiple 3D cuboids, moving continuously through near or far visual depths (see Methods). All subjects confirmed this stereoscopic percept during the scan session. All fMRI maps showed disparity-based near- and far-selective patches bilaterally in the inferior parietal cortex (Figure 3C and D).

Comparisons across maps sampled at different cortical depths suggested that these disparity-selective BOLD 'patches' in parietal cortex were radially elongated (i.e. columns), consistent with the radial elongation of disparity-selective columns shown in high field fMRI studies of human occipital cortex (Goncalves et al., 2015, Nasr et al., 2016, Tootell and Nasr, 2017, Nasr and Tootell, 2018, Nasr and Tootell, 2020). For simplicity, we refer to these disparity-sensitive parietal columns as 'D' (disparity) columns. In our parietal ROI, the near-selective disparity columns were slightly more numerous than the far-selective disparity columns, across the testable range of thresholds (above all thresholds between $10^{-2}$ through $10^{-5}$; t-test: all p-values < 0.01). This bias for near- (relative to far-) selective columns in parietal cortex may be related to a near-versus far-disparity bias reported in neighboring occipital cortex (Nasr and Tootell, 2020). In any event, we conclude that inferior parietal cortex includes visually driven, stereo-selective columns, similar to those previously described in occipital cortex.

Topographical relationship between column types

Next, we tested how the columns identified in Experiment 1 were topographically distributed relative to those in Experiment 3. Observations (Figures 2, 9A-D) suggested that the two categories of column (P and D), and/or the two functional poles in each of these categories (approaching versus withdrawing, and near versus far), might tend towards a *non-overlapping* arrangement in the cortical maps, analogous to the interdigitation of thin vs. thick type columns demonstrated in human visual cortical areas V2 and V3 (Nasr et al., 2016, Dumoulin et al., 2017, Tootell and Nasr, 2017). However, alternative topographic relationships are also possible, including: 1) a *random* organization of the two columnar categories and poles, or 2) a systematic *overlap* of the two columnar categories and/or poles (i.e. only one set of columns, which respond to all of our test stimuli, to varying extents).

By definition, two of the possible stimulus contrasts (near- vs. far-disparity, and approaching vs. withdrawing) are non-overlapping relative to each other, based on the nature of the fMRI subtraction *per se*. Thus, we tested only the relationship between maps evoked by the four remaining functional contrasts (near stereo vs. approach, far stereo vs. withdrawal, near stereo vs. withdrawal, and far stereo vs. withdrawal). BOLD response levels were measured in each vertex within each hemisphere, in response to these four functional contrasts, when the maps were: 1) accurately aligned (as *in vivo*), compared to 2) computationally 'shuffled' (pseudo-randomly misaligned).

The results confirmed that BOLD activity peaks in all four paired stimulus contrasts were preferentially non-overlapping (i.e. interdigitated) (Figure 9E-H). Based on these fMRI maps, one possible cortical 'unit' is illustrated in Figure 10.

## Discussion

Our evidence suggests that distinct fMRI responses to variations in stimulus distance are organized into at least two functional types of cortical columns within parietal cortex. Thus, one fundamental feature of the architecture of early visual cortex (i.e. functional segregation within multiple types of column) apparently also exists in parietal cortex.

Encoding of Interpersonal Distance

Broadly, our results suggest that specific columns in the parietal cortex map respond to variations in visually-mediated virtual distance from an observer, along either of two stimulus dimensions: 1) visual (i.e. binocular disparity), or 2) non-visual, apparently related to interpersonal distance. In both dimensions, the activity maps differed significantly in response to presentations of visual stimuli at 'near' versus 'far' extremes. In either dimension, future experiments conducted at an even finer spatial resolution might reveal a more continuous map including intermediate values.

However, the apparent congruence of 'near' and 'far' terms may be oversimplified, insofar as the two underlying dimensions differ from each other. For instance, psychophysical sensitivity to fused binocular disparity (stereopsis) ranges from ~ 10-20 cm from the eyes (depending on measurement technique) (reviewed in (Wilcox and Allison, 2009)), to a 'far' limit at optical inifinity. By comparison, sensitivity to personal space varies from a near limit at the observer's skin, to a far limit of ~ 50-100 cm (reviewed in (Hayduk, 1983)). The midpoint (zero crossing) distance values between the near-and-far extremes also differ. In the case of personal space, those zero-crossing values are unrelated to where the subject is looking. However, in the case of binocular disparity, the zero crossing distance varies enormously depending on where the subject is looking in each glance.

To optimize the experimental comprarison, here we studied both personal space and disparity by manipulating visual stimuli, centered along a common averaged line of sight. However, sensitivity to personal space is not limited to visual stimuli. For instance, prior studies (Hayduk, 1981b, Wormith, 1984, Bogovic et al., 2016, Hecht et al., 2019) suggest that personal space completely surrounds the body (albeit at different distances), i.e. in response to persons both visible and invisible. Auditory input (Lloyd et al., 2009, Noel et al., 2018) can also evoke PS-like discomfort when the source is interepreted as close. It has even been reported that proximity-driven discomfort can be evoked by simple statements that someone is within a subjects' personal space (Wormith, 1984, Kennedy et al., 2009), without any sensory cues. Consistent with this behavioral evidence, neurobiological studies have shown that sites in parietal cortex can be driven by multiple senses, in addition to vision (Andersen, 1997, Holmes and Spence, 2004, Murray and Wallace, 2011, Yang et al., 2011, Blanke, 2012, Ehrsson, 2012, Sereno and Huang, 2014, Grivaz et al., 2017).

Consistent with such generalization of cues for identifying personal space, the approach-selective columns showed stronger BOLD responses to progressively closer faces, when such faces were shown as *stationary* stimuli. Thus, based on their responses to both moving and stationary face stimuli, this category of columns could equally well be termed either 1) 'approach- or withdrawal-selective', or 2) 'interpersonally-near or far-selective' selective, respectively. In retrospect, our preferential use of the former term was largely arbitrary, reflecting the historical sequence in which we conducted our experiments.

Relationship to prior studies

In the current fMRI results, the disparity columns in parietal cortex were functionally similar to those described previously in visual areas V2, V3, V3A, and MT/V5 of macaque occipital cortex (Felleman and Van Essen, 1987, Poggio et al., 1988, Deangelis et al., 1998, Roy and Narahashi, 1992, Deangelis and Newsome, 1999, Adams and Zeki, 2001, Tanabe et al., 2005, Chen et al., 2008), and in likely-homologous areas in human visual cortex (Goncalves et al., 2015, Nasr et al., 2016, Parker et al., 2017, Nasr and Tootell, 2018, Nasr and Tootell, 2020). However, given the hierarchical nature of visual cue processing (e.g. (Livingstone and Hubel, 1988, Felleman and Van Essen, 1991, Ungerleider and Haxby, 1994, Ahissar and Hochstein, 2004, Siegle et al., 2021)) it would not be surprising if future experiments reveal novel aspects of disparity processing in the disparity responsive columns in parietal cortex, compared to those shown previously in occipital cortex.

It might be argued that our current results reflect a hemodynamic effect such as the radial 'diving veins' (Huber et al., 2015), unrelated to any columnar organization of neural responses. However, maps of metabolically active patches in macaque striate cortex (based on cytochrome oxidase staining) have not corresponded to the location of radially extending vessels (Keller et al., 2011, Adams et al., 2015). Furthermore, it is difficult to understand how multiple sets of functionally-distinct columns (e.g. a functional interdigitation, as found here and previously (Nasr et al., 2016, Dumoulin et al., 2017, Tootell and Nasr, 2017)) could arise from a single common vascular substrate.

Functionally, our column-scale fMRI findings in human parietal cortex are consistent with some single neuron results in macaque parietal cortex. For instance, a bias for approaching stimuli have been reported in single units in macaque parietal cortex (Colby et al., 1993, Graziano and Cooke, 2006), and in areas providing input to parietal cortex (Albright, 1989). Additional studies reported a preference for 'near' visual stimuli in monkey parietal cortex (Colby et al., 1993, Duhamel et al., 1998, Hadjidimitrakis et al., 2011, Bremmer et al., 2013, Cléry et al., 2017).

Within our ROI in parietal cortex, a discrete site (DPOS) showed especially strong BOLD responses to approaching and near faces. Previously, Quinlan and Culham (2007) identified human DPOS as an area that responded strongly to physically near (versus far) visual objects. However, those authors ultimately concluded that area DPOS responds selectively to inward vergence, rather than to near visual objects *per se*. However, that interpretation cannot fully explain the current results, in which the subjects' attention task required subjects to maintain a single vergence angle throughout the functional scans.

More than 25 sets of functional columns have been reported in multiple areas in primate visual cortex. However, previously there has been little evidence for analogous columns beyond visual cortex (but see (Bruce et al., 1985, Harvey et al., 2015). Insofar as the current approach/withdrawal columns encode interpersonal and/or social influence, the function of these parietal columns may be considered associative, rather than purely sensory.

Sensitive conventional fMRI techniques have described a coarse retinotopic organization in human parietal cortex (Arcaro et al., 2011, Swisher et al., 2007, Huang et al., 2017). However we could not identify any definite topographical relationship between the current results and such retinotopic variation.

As described above, parts of parietal cortex and cortical areas F4/5 can contribute to processing avoidance responses to approaching stimuli which are *physically* threatening. The current data emphasizes that this generality may be broadened to include stimuli that are *socially* threatening. In addition to parietal cortex and other threat-encoding areas, brain areas that might be engaged by threatening stimuli also include: 1) visual cortical sites involved in recognition of threatening objects, and 2) parietal and premotor cortical areas involved in sensorimotor preparation of defensive reactions to those threats.

<u>Clinical and public health implications</u>

Abnormalities in personal space regulation have been observed in multiple neuropsychiatric disorders, including schizophrenia (Holt et al., 2015, Horowitz et al., 1964, de la Asuncion et al., 2015, Nechamkin et al., 2003, Park et al., 2009, Srivastava and Mandal, 1990, Duke and Mullens, 1973) and autism (Gessaroli et al., 2013, Asada et al., 2016, Candini et al., 2017, Mul et al., 2019, Kennedy and Adolphs, 2014, Lough et al., 2015, Noel et al., 2017, Perry et al., 2015). Given our findings, it is possible that early changes in the organization of the parietal cortex and interconnected regions (perhaps including column-scale effects) are associated with abnormalities in personal space regulation observed in these neurodevelopmental conditions.

Although personal space preferences tend to be highly stable over time within individuals, some plasticity of personal space has also been observed. For example, tool use may lead to temporary expansion of personal space boundaries (Iriki et al., 1996, Berti and Frassinetti, 2000, Maravita and Iriki, 2004, Cardinali et al., 2009, Biggio et al., 2017, Iachini et al., 2014, Quesque et al., 2017) (but see (Patané et al., 2016)). Also, recent studies have suggested that preferred interpersonal distances have expanded during the COVID-19 pandemic, perhaps due to social distancing mandates - even when the actual risk of infection is minimal (Welsch et al., 2021, Iachini et al., 2020, Holt et al., 2021). It is not yet known whether such environmental factors affect personal space processing at the spatial scale of cortical columns.

# References

Adams, D. L., Piserchia, V., Economides, J. R. & Horton, J. C. 2015. Vascular supply of the cerebral cortex is specialized for cell layers but not columns. *Cerebral Cortex,* 25**,** 3673-3681.

Adams, D. L. & Zeki, S. 2001. Functional organization of macaque V3 for stereoscopic depth. *Journal of Neurophysiology,* 86**,** 2195-2203.

Adolphs, R. 2010. What does the amygdala contribute to social cognition? *Annals of the New York Academy of Sciences,* 1191**,** 42.

Ahissar, M. & Hochstein, S. 2004. The reverse hierarchy theory of visual perceptual learning. *Trends in cognitive sciences,* 8**,** 457-464.

Albright, T. 1989. Centrifugal directional bias in the middle temporal visual area (MT) of the macaque. *Visual neuroscience,* 2**,** 177-188.

Andersen, R. A. 1997. Multimodal integration for the representation of space in the posterior parietal cortex. *Philosophical Transactions of the Royal Society of London. Series B: Biological Sciences,* 352**,** 1421-1428.

Arcaro, M. J., Pinsk, M. A., Li, X. & Kastner, S. 2011. Visuotopic organization of macaque posterior parietal cortex: a functional magnetic resonance imaging study. *Journal of Neuroscience,* 31**,** 2064-2078.

Ardrey, R. 1966. The Territorial Imperative. New York: Atheneum.

Ashburner, J. & Friston, K. J. 2005. Unified segmentation. *Neuroimage,* 26**,** 839-851.

Bailenson, J. N., Blascovich, J., Beall, A. C. & Loomis, J. M. 2003. Interpersonal distance in immersive virtual environments. *Personality and social psychology bulletin,* 29**,** 819-833.

Bar-Haim, Y., Aviezer, O., Berson, Y. & Sagi, A. 2002. Attachment in infancy and personal space regulation in early adolescence. *Attachment & human development,* 4**,** 68-83.

Bastiani, M., Cottaar, M., Fitzgibbon, S. P., Suri, S., Alfaro-Almagro, F., Sotiropoulos, S. N., Jbabdi, S. & Andersson, J. L. 2019. Automated quality control for within and between studies diffusion MRI data using a non-parametric framework for movement and distortion correction. *Neuroimage,* 184**,** 801-812.

Blanke, O. 2012. Multisensory brain mechanisms of bodily self-consciousness. *Nature Reviews Neuroscience,* 13**,** 556-571.

Bogdanova, O. V., Bogdanov, V. B., Dureux, A., Farnè, A. & Hadj-Bouziane, F. 2021. The Peripersonal Space in a social world. *Cortex,* 142**,** 28-46.

Bogovic, A., Ivezic, E. & Filipcic, I. 2016.

Personal space of war veterans with PTSD-SOME characteristics and comparison with healthy individuals. *Psychiatria Danubina,* 28**,** 0-81.

Brainard, D. H. 1997. The psychophysics toolbox. *Spatial vision,* 10**,** 433-436.

Bremmer, F., Schlack, A., Kaminiarz, A. & Hoffmann, K. P. 2013. Encoding of movement in near extrapersonal space in primate area VIP. *Frontiers in behavioral neuroscience,* 7**,** 8.

Bruce, C. J., Goldberg, M. E., Bushnell, M. C. & Stanton, G. 1985. Primate frontal eye fields. II. Physiological and anatomical correlates of electrically evoked eye movements. *Journal of neurophysiology,* 54**,** 714-734.

Candini, M., Battaglia, S., Benassi, M., Di Pellegrino, G. & Frassinetti, F. 2021. The physiological correlates of interpersonal space. *Scientific Reports,* 11**,** 1-8.

Cartaud, A., Quesque, F. & Coello, Y. 2020. Wearing a face mask against Covid-19 results in a reduction of social distancing. *Plos one,* 15.

Chen, G., Lu, H. D. & Roe, A. W. 2008. A map for horizontal disparity in monkey V2. *Neuron,* 58**,** 442-450.

Cléry, J., Guipponi, O., Odouard, S., Pinède, S., Wardak, C. & Hamed, S. B. 2017. The prediction of impact of a looming stimulus onto the body is subserved by multisensory integration mechanisms. *Journal of Neuroscience,* 37**,** 10656-10670.

Cléry, J., Guipponi, O., Odouard, S., Wardak, C. & Hamed, S. B. 2015. Impact prediction by looming visual stimuli enhances tactile detection. *Journal of Neuroscience,* 35**,** 4179-4189.

Cléry, J. & Hamed, S. B. 2018. Frontier of self and impact prediction. *Frontiers in psychology,* 9**,** 1073.

Colby, C. L., Duhamel, J.-R. & Goldberg, M. E. 1993. Ventral intraparietal area of the macaque: anatomic location and visual response properties. *Journal of neurophysiology,* 69**,** 902-914.

Cooke, D. F., Taylor, C. S., Moore, T. & Graziano, M. S. 2003. Complex movements evoked by microstimulation of the ventral intraparietal area. *Proceedings of the National Academy of Sciences,* 100**,** 6163-6168.

Dale, A. M., Fischl, B. & Sereno, M. I. 1999. Cortical surface-based analysis: I. Segmentation and surface reconstruction. *Neuroimage,* 9**,** 179-194.

Dale, A. M. & Sereno, M. I. 1993. Improved localizadon of cortical activity by combining EEG and MEG with MRI cortical surface reconstruction: a linear approach. *Journal of cognitive neuroscience,* 5**,** 162-176.

Deangelis, G. C., Cumming, B. G. & Newsome, W. T. 1998. Cortical area MT and the perception of stereoscopic depth. *Nature,* 394**,** 677-680.

Deangelis, G. C. & Newsome, W. T. 1999. Organization of disparity-selective neurons in macaque area MT. *Journal of Neuroscience,* 19**,** 1398-1415.

Duhamel, J.-R., Colby, C. L. & Goldberg, M. E. 1998. Ventral intraparietal area of the macaque: congruent visual and somatic response properties. *Journal of neurophysiology,* 79**,** 126-136.

Dumoulin, S. O., Harvey, B. M., Fracasso, A., Zuiderbaan, W., Luijten, P. R., Wandell, B. A. & Petridou, N. 2017. In vivo evidence of functional and anatomical stripe-based subdivisions in human V2 and V3. *Scientific reports,* 7**,** 1-12.

Ehrsson, H. H. 2012. 43 The concept of body ownership and its relation to multisensory integration. *The New Handbook of Multisensory Process*.

Emery, N. J., Capitanio, J. P., Mason, W. A., Machado, C. J., Mendoza, S. P. & Amaral, D. G. 2001. The effects of bilateral lesions of the amygdala on dyadic social interactions in rhesus monkeys (Macaca mulatta). *Behavioral neuroscience,* 115**,** 515.

Felleman, D. J. & Van Essen, D. C. 1987. Receptive field properties of neurons in area V3 of macaque monkey extrastriate cortex. *Journal of neurophysiology,* 57**,** 889-920.

Felleman, D. J. & Van Essen, D. C. 1991. Distributed hierarchical processing in the primate cerebral cortex. *Cerebral cortex (New York, NY: 1991),* 1**,** 1-47.

Fischl, B. & Dale, A. M. 2000. Measuring the thickness of the human cerebral cortex from magnetic resonance images. *Proceedings of the National Academy of Sciences,* 97**,** 11050-11055.

Fischl, B., Liu, A. & Dale, A. M. 2001. Automated manifold surgery: constructing geometrically accurate and topologically correct models of the human cerebral cortex. *IEEE transactions on medical imaging,* 20**,** 70-80.

Fischl, B., Salat, D. H., Busa, E., Albert, M., Dieterich, M., Haselgrove, C., Van Der Kouwe, A., Killiany, R., Kennedy, D. & Klaveness, S. 2002. Whole brain segmentation: automated labeling of neuroanatomical structures in the human brain. *Neuron,* 33**,** 341-355.

Fischl, B., Salat, D. H., Van Der Kouwe, A. J., Makris, N., Ségonne, F., Quinn, B. T. & Dale, A. M. 2004a. Sequence-independent segmentation of magnetic resonance images. *Neuroimage,* 23**,** S69-S84.

Fischl, B., Sereno, M. I., Tootell, R. B. & Dale, A. M. 1999. High-resolution intersubject averaging and a coordinate system for the cortical surface. *Human brain mapping,* 8**,** 272-284.

Fischl, B., Van Der Kouwe, A., Destrieux, C., Halgren, E., Ségonne, F., Salat, D. H., Busa, E., Seidman, L. J., Goldstein, J. & Kennedy, D. 2004b. Automatically parcellating the human cerebral cortex. *Cerebral cortex,* 14**,** 11-22.

Goncalves, N. R., Ban, H., Sánchez-Panchuelo, R. M., Francis, S. T., Schluppeck, D. & Welchman, A. E. 2015. 7 tesla FMRI reveals systematic functional organization for binocular disparity in dorsal visual cortex. *Journal of Neuroscience,* 35**,** 3056-3072.

Graziano, M. S. & Cooke, D. F. 2006. Parieto-frontal interactions, personal space, and defensive behavior. *Neuropsychologia,* 44**,** 845-859.

Graziano, M. S., Taylor, C. S. & Moore, T. 2002. Complex movements evoked by microstimulation of precentral cortex. *Neuron,* 34**,** 841-851.

Greve, D. N. & Fischl, B. 2009. Accurate and robust brain image alignment using boundary-based registration. *Neuroimage,* 48**,** 63-72.

Grivaz, P., Blanke, O. & Serino, A. 2017. Common and distinct brain regions processing multisensory bodily signals for peripersonal space and body ownership. *Neuroimage,* 147**,** 602-618.

Hadjidimitrakis, K., Bertozzi, F., Breveglieri, R., Fattori, P. & Galletti, C. 2014. Body-centered, mixed, but not hand-centered coding of visual targets in the medial posterior parietal cortex during reaches in 3D space. *Cerebral Cortex,* 24**,** 3209-3220.

Hadjidimitrakis, K., Breveglieri, R., Placenti, G., Bosco, A., Sabatini, S. P. & Fattori, P. 2011. Fix your eyes in the space you could reach: neurons in the macaque medial parietal cortex prefer gaze positions in peripersonal space. *PLoS One,* 6**,** e23335.

Hall, E. T. 1966. *The hidden dimension*, Garden City, NY: Doubleday.

Harvey, B. M., Fracasso, A., Petridou, N. & Dumoulin, S. O. 2015. Topographic representations of object size and relationships with numerosity reveal generalized quantity processing in human parietal cortex. *Proceedings of the National Academy of Sciences,* 112**,** 13525-13530.

Hayduk, L. A. 1978. Personal space: An evaluative and orienting overview. *Psychological bulletin,* 85**,** 117.

Hayduk, L. A. 1981a. The permeability of personal space. *Canadian Journal of Behavioural Science/Revue canadienne des sciences du comportement,* 13**,** 274.

Hayduk, L. A. 1981b. The shape of personal space: An experimental investigation. *Canadian Journal of Behavioural Science/Revue canadienne des sciences du comportement,* 13**,** 87.

Hayduk, L. A. 1983. Personal Space: Where We Now Stand. *Psychological Bulletin,* 94**,** 293-335.

Hecht, H., Welsch, R., Viehoff, J. & Longo, M. R. 2019. The shape of personal space. *Acta psychologica,* 193**,** 113-122.

Hediger, H. 1955. Studies of the psychology and behavior of captive animals in zoos and circuses.

Holmes, N. P. & Spence, C. 2004. The body schema and multisensory representation (s) of peripersonal space. *Cognitive processing,* 5**,** 94-105.

Holt, D. J., Cassidy, B. S., Yue, X., Rauch, S. L., Boeke, E. A., Nasr, S., Tootell, R. B. & Coombs, G. 2014. Neural correlates of personal space intrusion. *Journal of Neuroscience,* 34**,** 4123-4134.

Holt, D. J., Zapetis, S., Babadi, B. & Tootell, R. B. H. 2021. Personal space Increases during the COVID-19 Pandemic in Response to Real and Virtual Humans. *medRxiv***,** 2021.06.09.21258234.

Horton, J. C. & Adams, D. L. 2005. The cortical column: a structure without a function. *Philosophical Transactions of the Royal Society B: Biological Sciences,* 360**,** 837-862.

Horton, J. C. & Hocking, D. R. 1996. Intrinsic variability of ocular dominance column periodicity in normal macaque monkeys. *Journal of Neuroscience,* 16**,** 7228-7339.

Huang, R.-S., Chen, C.-F. & Sereno, M. I. 2017. Mapping the complex topological organization of the human parietal face area. *NeuroImage,* 163**,** 459-470.

Hubel, D. H. & Wiesel, T. N. 1962. Receptive fields, binocular interaction and functional architecture in the cat's visual cortex. *The Journal of physiology,* 160**,** 106-154.

Hubel, D. H. & Wiesel, T. N. 1968. Receptive fields and functional architecture of monkey striate cortex. *The Journal of physiology,* 195**,** 215-243.

Huber, L., Goense, J., Kennerley, A. J., Trampel, R., Guidi, M., Reimer, E., Ivanov, D., Neef, N., Gauthier, C. J. & Turner, R. 2015. Cortical lamina-dependent blood volume changes in human brain at 7 T. *Neuroimage,* 107**,** 23-33.

Iachini, T., Coello, Y., Frassinetti, F. & Ruggiero, G. 2014. Body space in social interactions: a comparison of reaching and comfort distance in immersive virtual reality. *PloS one,* 9**,** e111511.

Iachini, T., Coello, Y., Frassinetti, F., Senese, V. P., Galante, F. & Ruggiero, G. 2016. Peripersonal and interpersonal space in virtual and real environments: Effects of gender and age. *Journal of Environmental Psychology,* 45**,** 154-164.

Iachini, T., Frassinetti, F., Ruotolo, F., Sbordone, F. L., Ferrara, A., Arioli, M., Pazzaglia, F., Bosco, A., Candini, M. & Lopez, A. 2020. Psychological and situational effects on social distancing and well-being during the COVID-19 pandemic: not a question of real risk.

Keller, A. L., Schüz, A., Logothetis, N. K. & Weber, B. 2011. Vascularization of cytochrome oxidase-rich blobs in the primary visual cortex of squirrel and macaque monkeys. *Journal of Neuroscience,* 31**,** 1246-1253.

Kennedy, D. P., Gläscher, J., Tyszka, J. M. & Adolphs, R. 2009. Personal space regulation by the human amygdala. *Nature neuroscience,* 12**,** 1226-1227.

Klüver, H. & Bucy, P. C. 1937. " Psychic blindness" and other symptoms following bilateral temporal lobectomy in Rhesus monkeys. *American Journal of Physiology*.

Livingstone, M. & Hubel, D. 1988. Segregation of form, color, movement, and depth: anatomy, physiology, and perception. *Science,* 240**,** 740-749.

Llobera, J., Spanlang, B., Ruffini, G. & Slater, M. 2010. Proxemics with multiple dynamic characters in an immersive virtual environment. *ACM Transactions on Applied Perception (TAP),* 8**,** 1-12.

Lloyd, D., Morrison, I. & Roberts, N. 2006. Role for human posterior parietal cortex in visual processing of aversive objects in peripersonal space. *Journal of neurophysiology,* 95**,** 205-214.

Lloyd, D. M., Coates, A., Knopp, J., Oram, S. & Rowbotham, S. 2009. Don't stand so close to me: The effect of auditory input on interpersonal space. *Perception,* 38**,** 617-620.

Logothetis, N. K. 2003. The underpinnings of the BOLD functional magnetic resonance imaging signal. *Journal of Neuroscience,* 23**,** 3963-3971.

Logothetis, N. K., Pauls, J., Augath, M., Trinath, T. & Oeltermann, A. 2001. Neurophysiological investigation of the basis of the fMRI signal. *nature,* 412**,** 150-157.

Luppino, G., Murata, A., Govoni, P. & Matelli, M. 1999. Largely segregated parietofrontal connections linking rostral intraparietal cortex (areas AIP and VIP) and the ventral premotor cortex (areas F5 and F4). *Experimental Brain Research,* 128**,** 181-187.

Mason, W. A., Capitanio, J. P., Machado, C. J., Mendoza, S. P. & Amaral, D. G. 2006. Amygdalectomy and responsiveness to novelty in rhesus monkeys (Macaca mulatta): generality and individual consistency of effects. *Emotion,* 6**,** 73.

Mccall, C., Blascovich, J., Young, A. & Persky, S. 2009. Proxemic behaviors as predictors of aggression towards Black (but not White) males in an immersive virtual environment. *Social Influence,* 4**,** 138-154.

Mountcastle, V. B. 1957. Modality and topographic properties of single neurons of cat's somatic sensory cortex. *Journal of neurophysiology,* 20**,** 408-434.

Murray, M. M. & Wallace, M. T. 2011. The neural bases of multisensory processes.

Nacewicz, B. M., Dalton, K. M., Johnstone, T., Long, M. T., Mcauliff, E. M., Oakes, T. R., Alexander, A. L. & Davidson, R. J. 2006. Amygdala volume and nonverbal social impairment in adolescent and adult males with autism. *Archives of general psychiatry,* 63**,** 1417-1428.

Nasr, S., Polimeni, J. R. & Tootell, R. B. 2016. Interdigitated color-and disparity-selective columns within human visual cortical areas V2 and V3. *Journal of Neuroscience,* 36**,** 1841-1857.

Nasr, S. & Tootell, R. B. 2018. Visual field biases for near and far stimuli in disparity selective columns in human visual cortex. *Neuroimage,* 168**,** 358-365.

Nasr, S. & Tootell, R. B. 2020. Asymmetries in Global Perception Are Represented in Near-versus Far-Preferring Clusters in Human Visual Cortex. *Journal of Neuroscience,* 40**,** 355-368.

Noel, J.-P., Samad, M., Doxon, A., Clark, J., Keller, S. & Di Luca, M. 2018. Peri-personal space as a prior in coupling visual and proprioceptive signals. *Scientific reports,* 8**,** 1-15.

Parker, A. J., Coullon, G., Sanchez-Panchuelo, R., Francis, S., Clare, S., Kay, D., Duff, E., Minini, L., Jbabdi, S. & Schluppeck, D. 2017. Geospatial statistics of high field functional MRI reveals topographical clustering for binocular stereo depth in early visual cortex. *bioRxiv***,** 160788.

Pelli, D. G. 1997. The VideoToolbox software for visual psychophysics: Transforming numbers into movies. *Spatial vision,* 10**,** 437-442.

Poggio, G. F., Gonzalez, F. & Krause, F. 1988. Stereoscopic mechanisms in monkey visual cortex: binocular correlation and disparity selectivity. *Journal of Neuroscience,* 8**,** 4531-4550.

Polimeni, J. R., Fischl, B., Greve, D. N. & Wald, L. L. 2010. Laminar analysis of 7 T BOLD using an imposed spatial activation pattern in human V1. *Neuroimage,* 52**,** 1334-1346.

Power, J. D., Barnes, K. A., Snyder, A. Z., Schlaggar, B. L. & Petersen, S. E. 2012. Spurious but systematic correlations in functional connectivity MRI networks arise from subject motion. *Neuroimage,* 59**,** 2142-2154.

Quinlan, D. & Culham, J. C. 2007. fMRI reveals a preference for near viewing in the human parieto-occipital cortex. *Neuroimage,* 36**,** 167-187.

Rinck, M., Rörtgen, T., Lange, W.-G., Dotsch, R., Wigboldus, D. H. & Becker, E. S. 2010. Social anxiety predicts avoidance behaviour in virtual encounters. *Cognition and Emotion,* 24**,** 1269-1276.

Roy, M. L. & Narahashi, T. 1992. Differential properties of tetrodotoxin-sensitive and tetrodotoxin-resistant sodium channels in rat dorsal root ganglion neurons. *Journal of Neuroscience,* 12**,** 2104-2111.

Ruggiero, G., D'errico, O. & Iachini, T. 2016. Development of egocentric and allocentric spatial representations from childhood to elderly age. *Psychological research,* 80**,** 259-272.

Schienle, A., Wabnegger, A., Schöngassner, F. & Leutgeb, V. 2015. Effects of personal space intrusion in affective contexts: an fMRI investigation with women suffering from borderline personality disorder. *Social cognitive and affective neuroscience,* 10**,** 1424-1428.

Sereno, M. I. & Huang, R.-S. 2014. Multisensory maps in parietal cortex. *Current opinion in neurobiology,* 24**,** 39-46.

Siegle, J. H., Jia, X., Durand, S., Gale, S., Bennett, C., Graddis, N., Heller, G., Ramirez, T. K., Choi, H. & Luviano, J. A. 2021. Survey of spiking in the mouse visual system reveals functional hierarchy. *Nature,* 592**,** 86-92.

Smyser, C. D., Inder, T. E., Shimony, J. S., Hill, J. E., Degnan, A. J., Snyder, A. Z. & Neil, J. J. 2010. Longitudinal analysis of neural network development in preterm infants. *Cerebral cortex,* 20**,** 2852-2862.

Swisher, J. D., Halko, M. A., Merabet, L. B., Mcmains, S. A. & Somers, D. C. 2007. Visual topography of human intraparietal sulcus. *Journal of Neuroscience,* 27**,** 5326-5337.

Tanabe, S., Doi, T., Umeda, K. & Fujita, I. 2005. Disparity-tuning characteristics of neuronal responses to dynamic random-dot stereograms in macaque visual area V4. *Journal of neurophysiology,* 94**,** 2683-2699.

Todd, R. M. & Anderson, A. K. 2009. Six degrees of separation: the amygdala regulates social behavior and perception. *Nature neuroscience,* 12**,** 1217-1218.

Tootell, R., Hamilton, S., Silverman, M. & Switkes, E. 1988. Functional anatomy of macaque striate cortex. I. Ocular dominance, binocular interactions, and baseline conditions. *Journal of Neuroscience,* 8**,** 1500-1530.

Tootell, R. B. & Nasr, S. 2017. Columnar segregation of magnocellular and parvocellular streams in human extrastriate cortex. *Journal of Neuroscience,* 37**,** 8014-8032.

Tootell, R. B., Zapetis, S., Babadi, B., Nasiriavanaki, Z., Hughes, D., Mueser, K., Otto, M., Pace-Schott, E. & Holt, D. J. 2021. Psychological and Physiological Evidence for an Initial 'Rough Sketch' Calculation of Personal Space. *Scientific Reports*.

Tsao, D. Y., Vanduffel, W., Sasaki, Y., Fize, D., Knutsen, T. A., Mandeville, J. B., Wald, L. L., Dale, A. M., Rosen, B. R. & Van Essen, D. C. 2003. Stereopsis activates V3A and caudal intraparietal areas in macaques and humans. *Neuron,* 39**,** 555-568.

Ungerleider, L. G. & Haxby, J. V. 1994. 'What'and 'where'in the human brain. *Current opinion in neurobiology,* 4**,** 157-165.

Wabnegger, A., Leutgeb, V. & Schienle, A. 2016. Differential amygdala activation during simulated personal space intrusion by men and women. *Neuroscience,* 330**,** 12-16.

Welsch, R., Hecht, H., Chuang, L. & Von Castell, C. 2020. Interpersonal distance in the SARS-CoV-2 Crisis. *Human Factors,* 62**,** 1095-1101.

Welsch, R., Von Castell, C. & Hecht, H. 2019. The anisotropy of personal space. *PloS one,* 14**,** e0217587.

Welsch, R., Wessels, M., Bernhard, C., Thönes, S. & Von Castell, C. 2021. Physical distancing and the perception of interpersonal distance in the COVID-19 crisis. *Scientific reports,* 11**,** 1-9.

Whalen, P. J., Rauch, S. L., Etcoff, N. L., Mcinerney, S. C., Lee, M. B. & Jenike, M. A. 1998. Masked presentations of emotional facial expressions modulate amygdala activity without explicit knowledge. *Journal of neuroscience,* 18**,** 411-418.

Wieser, M. J., Pauli, P., Grosseibl, M., Molzow, I. & Mühlberger, A. 2010. Virtual social interactions in social anxiety—the impact of sex, gaze, and interpersonal distance. *Cyberpsychology, Behavior, and Social Networking,* 13**,** 547-554.

Wilcox, L. M. & Allison, R. S. 2009. Coarse-fine dichotomies in human stereopsis. *Vision research,* 49**,** 2653-2665.

Williams, J. L. 1971. Personal space and its relation to extraversion-introversion. *Canadian Journal of Behavioural Science/Revue canadienne des sciences du comportement,* 3**,** 156.

Wilson, H. R. & Cowan, J. D. 1972. Excitatory and inhibitory interactions in localized populations of model neurons. *Biophysical journal,* 12**,** 1-24.

Wormith, J. 1984. Personal space of incarcerated offenders. *Journal of clinical psychology,* 40**,** 815-827.

Yang, Y., Liu, S., Chowdhury, S. A., Deangelis, G. C. & Angelaki, D. E. 2011. Binocular disparity tuning and visual–vestibular congruency of multisensory neurons in macaque parietal cortex. *Journal of Neuroscience,* 31**,** 17905-17916.

Zaretskaya, N., Fischl, B., Reuter, M., Renvall, V. & Polimeni, J. R. 2018. Advantages of cortical surface reconstruction using submillimeter 7 T MEMPRAGE. *Neuroimage,* 165**,** 11-26.

**Figure Titles and Legends**

**Figure 1**. **Personal space measurements differ between subjects, but remain relatively consistent across repeated measurements within a subject**. Panel A shows measurements acquired outside the scanner, using the standard passive stop distance procedure, in response to real human confederates ('intruders'). Panel B shows analogous measurements to intruding face images, acquired while subjects were in the 7T scanner, but without MR imaging. Following a practice trial, personal space was measured in 5 (top panel) or 15 (bottom panel) consecutive trials, in each subject (n = 7) who was subsequently scanned. In both panels, data from each subject are shown in a different, arbitrarily assigned color (upper right). The means are superimposed as a dashed white line. Error bars show one standard error of the mean.

**Figure 2. Group averaged maps, approach versus withdrawal.** Panels A and B show the group-averaged cortical activity (n=7) in response to virtually approaching (compared to withdrawing) face images (7T, 1.1 mm iso). Data are shown from both hemispheres ("L" and "R" at the top, for "left" and "right"), in the corresponding cortical surfaces, viewed from a posterior-dorsal viewpoint in inflated cortical format (Panel A), and in a cortically flattened view (Panel B). The activity scale on the bottom left indicates -log10(p), where p is chance probability (yellow/red = approach > withdrawal; cyan/blue = withdrawal > approach). Our region-of-interest (within thin white lines, in posterior parietal cortex) is based on the gyral/sulcal anatomy in the group-averaged cortical surface. In Panel B, the white arrows indicate the dorsal posterior occipital sulcus (DPOS). The spatial scale bar (bottom right) is applicable to the flattened view, subject to distortion (+/- 15%) due to cortical flattening.

**Figure 3. High resolution activity maps from single subjects.** All panels show flattened maps of cortical activity in both hemispheres in individual subjects (#1 and 7). Anterior/posterior is upwards/downwards in the figure, respectively. Medial is towards the midline between Panels A and B. The parietal region of interest is bounded by thin solid white lines. Panel A: Approach- versus withdrawal-selective activity is shown in red-yellow and blue-cyan, respectively (activity scale bar for all panels on the right). Panels C and D show activity maps of near- and far-selective disparity columns (red-yellow and blue-cyan, respectively), in response to visual presentation of random dot stereograms (see below). Data in panels C and D are from the same subjects as Panels A and B (subjects #1 and 7), to facilitate comparison of the two maps. (see below). In panels C and D, functionally similar disparity columns are also evident within visual cortical area V3A (in and near green Xs), as described previously.

**Figure 4. Consistency of evoked activity maps across sessions.** Left panel: Consistency of BOLD activity maps in flattened cortex from our ROI, in response to the approaching versus withdrawing stimuli, across each of the two sessions (panels A-D), and their average across both sessions (panels E, F), from subjects 7 and 1 (left and right panels, respectively). In panels A-B, yellow/red = approach > withdrawal; cyan/blue = withdrawal > approach. Right panel: An overlap analysis confirms the consistency of the within-subject maps across scanning sessions. Panel G shows the group-averaged (n=7) level of overlap in within-subject maps of approach > withdrawal activity, when generated in session 1 compared to session 2, when the two maps were aligned correctly (left bar, red), and when pseudo-randomly mis-aligned ('shuffled', averaged over 1000 iterations) (right bar, light red) (t-test: $p < 0.0001$). Panel H shows the analogous levels of overlap in maps of withdrawal > approach activity, when aligned (left bar, blue) and when mis-aligned (right bar, light blue) (t-test: $p < 0.0001$). Brackets represent one standard error of the mean.

**Figure 5. The patchy topography in the approach vs. withdrawal maps remained very similar across variations in cortical depth**. Maps of approach vs. withdrawal responses of right and left hemispheres of subject 1 are shown across variations in cortical depth. Panel A shows the activity map sampled from vertices centered on the cortical surface (depth 0), which included activity in cortical layers 1 and 2. The activity in Panel B was (on average) centered midway through the gray matter (depth 0.5), including cortical

layer 4. The activity map in Panel C was centered at a depth of 0.8, including cortical layers 5 and 6. Panel D was located at the gray/white matter boundary (depth 1.0), including activity in layer 6. The amplitude in panel D was reduced due to partial volume inclusion of the white matter, which shows negligible BOLD variation.

**Figure 6. Activity evoked by the approaching and withdrawing face stimuli is radially elongated (columnar)**. Flattened activity maps in the parietal region of interest were sampled at 20 equally spaced cortical depths, which were then radially aligned and combined into a laminar 'sandwich'. Activity was sampled and averaged above and below a line drawn across the cortical surface (white line, Panels A-D). Panels C and D show the same data as in panel A and B, except that values in the activity map were re-scaled to match the broader range of pseudo-color activity scaling in panels E and F, in views analogous to classical views of cortical columns. Panels G-J show additional examples of the BOLD activity across depth from other cortical locations (not shown), otherwise as in panels E and F. In all panels, the BOLD activity was elongated along the radial axis (i.e. vertical). Given the averaged gray matter width in this region (2.37 mm), and the voxel size (1.1 mm$^3$, iso), the activity maps were nominally independent (not vertically overlapping) at the two depth extremes (0 and 1). Activity at other depths could partially overlap, e.g. reflecting the relative depth levels and the local cortical thickness. Panels A, C and E are illustrated from subject 5 (left hemisphere, thickness within our ROI = 2.21 mm). Panels B, D and F are from subject 2 (right hemisphere, thickness = 2.46 mm). Panel G = subject 5 (right hemisphere, thickness = 2.28 mm); panel H = subject 5 (left hemisphere, thickness = 2.21 mm); Panel I = subject 1 (right hemisphere, thickness = 2.15 mm); Panel J = subject 7 (right hemisphere, thickness = 2.41 mm). Yellow/red = approach > withdrawal; cyan/blue = withdrawal > approach.

**Figure 7. Correlation of functional activity along radial- versus surface-parallel axes.** Correlation scatterplots of evoked approach- vs. withdrawal-biased BOLD levels sampled along radial vs. surface-parallel axes in the flattened cortical surfaces within our ROI, averaged across all (n = 14) hemispheres.

The sampling planes are schematized in panels A and C. Panel B shows the correlation of activity in vertices when sampled along an axis perpendicular (radial) to the cortical surface (red lines). Panels D and E show the correlation values when sampled across a comparable distance in the perpendicular cortical plane, i.e. parallel with the cortical surfaces and layers. The sampling plane was deep (depth = 0.9) in panel D, and more superficial (depth = 0.1) in panel E. Thus, the functional response variations were more highly correlated in the radial axis, compared to the surface-parallel axis.

**Figure 8. BOLD responses to variations in stationary distance.** Panel A shows examples of the nine stimulus images used in the experiment. The face stimuli in panel A are shown from left to right in order of increasing proximity (decreasing interpersonal distance), but in the actual experiment, the presentation order was pseudo-randomized. Panel B shows the average BOLD peak amplitude in response to the variations in such stationary face stimuli, when presented at different virtual distances from the subject. The response functions were measured independently in the approach-selective columns (red), and in the withdrawal-selecctive columns (cyan). The vertical dashed white line indicates the response at each subject's personal space boundary (100%); other distances were normalized, pre-calculated and presented as percentages of that value, for each subject (see Methods).

Panel C shows one possible scheme to compute such parietal responses (bottom) from visual cortical inputs (top). Results from Experiments 1 and 2 suggested that the input corresponding to physically closer faces was larger in size, hence stronger. The average firing rates of the excitatory and inhibitory populations in inferior parietal cortex were represented by rate equations (Wilson and Cowan, 1972), in which each population received weighted inputs from the other population and itself, through a sigmoid nonlinearity (Table 1). For simplicity, this model assumed that BOLD levels represent the excitatory postsynaptic response, although it is known that BOLD signals reflect a combination of both pre- and post-synaptic activity (Logothetis et al., 2001, Logothetis, 2003). The synaptic weights between and within the two populations were considered free parameters, which were tuned to approximate the response profiles of approach- and withdrawal-selective columns (Table 1).

In the approach-selective columns, the model suggested that a strong recurrent excitation and a weak recurrent inhibition could account for the activity profile of the excitatory population in response to the upstream visual input (Panel C, top, orange and purple respectively). Due to the weak recurrent inhibition, the response of the excitatory population increased in response to the input. When the input was strong enough to activate the excitatory population beyond its intrinsic soft threshold, there was a rapid increase in the excitatory firing rate due to the strong recurrent excitation (Panel C, bottom, red). In the withdrawal-selective columns, a weak recurrent excitation and a strong recurrent inhibition could account for the response of the excitatory population to the input. For weaker inputs, inhibition dominated over excitation; thus the excitatory firing decreased as a function of the input. However, as the input became even stronger, the activity of the inhibitory population saturated and excitation overcame the inhibition, which eventually caused the excitatory firing rate to plateau (Panel C, bottom, cyan).

**Figure 9. Topographic distribution of presumptive personal and disparity-based columns.** Panels A-D (leftmost) show maps of the cortical overlap (and non-overlap) in the flattened activity maps of approach- and withdrawal-biased activity, plus the activity in response to near- and far-biased binocular disparity, as in a combination of the individual activity maps shown in Figure 3. Here, analogous maps were recolorized and overlapped (here, left and right hemispheres from subject 7 (panels A, B) and subject 2 (panels C, D), within our ROI (white boundary lines). In those four maps, vertices that showed significantly biased activity to any single one (but not two or more) of the four functional contrasts are color-coded as shown below panels C and D. Vertices that showed overlap (i.e. to two or more) of any of these functional contrasts are shown in black. The significance threshold for all contrasts was $p < 0.05$, uncorrected. Observations suggest that there was relatively little overlap, implying a segregation of function. Panels E-H (rightmost) show the extent of overlap in the maps of the leftmost panels, when averaged across all hemispheres (n = 14), within our parietal ROI, for all four pairwise contrasts of interest, across a range of activity thresholds (x axis). The magenta line shows the extent of overlap when the maps were correctly aligned (as *in vivo*), across different thresholds (x axis). As a control condition, the black line shows the level of overlap when the maps were aligned pseudo-randomly relative to each other, averaged across 1000 randomizations.

Overlap is defined as the percentage of overlapped vertices (i. e. the number of vertices showing selectivity for a given functional contrast, divided by the number of vertices within our ROI). The results confirm a strong tendency towards interdigitation (e.g. mutual repulsion) between all four of the functional contrasts.

**Figure 10. Hypothetical column-scale distance-processing unit in parietal cortex.** The cortical surface is drawn at the top of the figure. DN = Near disparity; DF = Far disparity; PN = Near personal; PF = Far personal. The diagram is based on fMRI data (e.g. Figure 9).

**Table 1. Parameters of the computational model.** Numerical values of the parameters in the current scheme. Following the widely used framework of the firing rate neural models (Wilson and Cowan 1972), the firing rates of excitatory and inhibitory populations were modeled as:

$$\frac{dr_E}{dt} = -r_E + W_{EE}\, f(r_E) + W_{EI}\, f(r_I) + \theta_E + W_E\, I$$

$$\frac{dr_I}{dt} = -r_I + W_{IE}\, f(r_E) + W_{II}\, f(r_I) + \theta_I + W_I\, I$$

where $r_E$ and $r_I$ are the firing rate of excitatory and inhibitory populations, respectively. $W_{EE}$ is the synaptic strength of excitatory-to-excitatory connections, $W_{EI}$ is the strength of inhibitory-to-excitatory connections, $W_{IE}$ is the strength of excitatory-to-inhibitory connections, and $W_{II}$ is the strength of inhibitory-to-inhibitory connections. Each population has a baseline activity represented by $\theta_E$ and $\theta_I$, respectively. Finally, each population receives the external input $I$ weighted by synaptic strengths $W_E$ and $W_I$, respectively. Each population receives the firing rate of the other population as input through a sigmoid nonlinear function $f(x) = 1/(1 + \exp(-(x + \theta)))$, where $\theta$ represents the intrinsic soft threshold of the population activity. The variable of interest in our simulations was the stable steady-state firing rate of excitatory population as a function of the strength of the external input $I$.

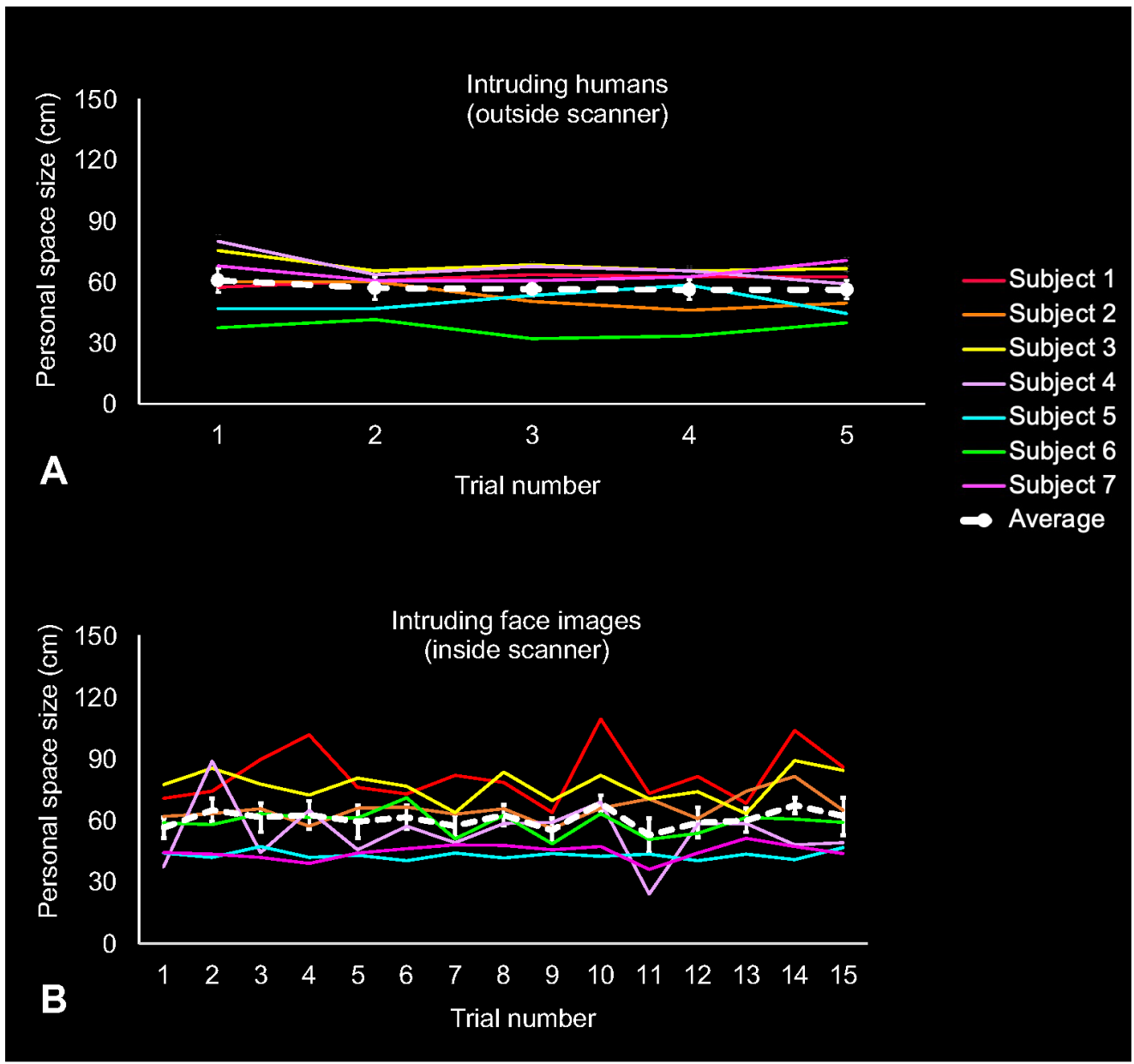

**Figure 1.**

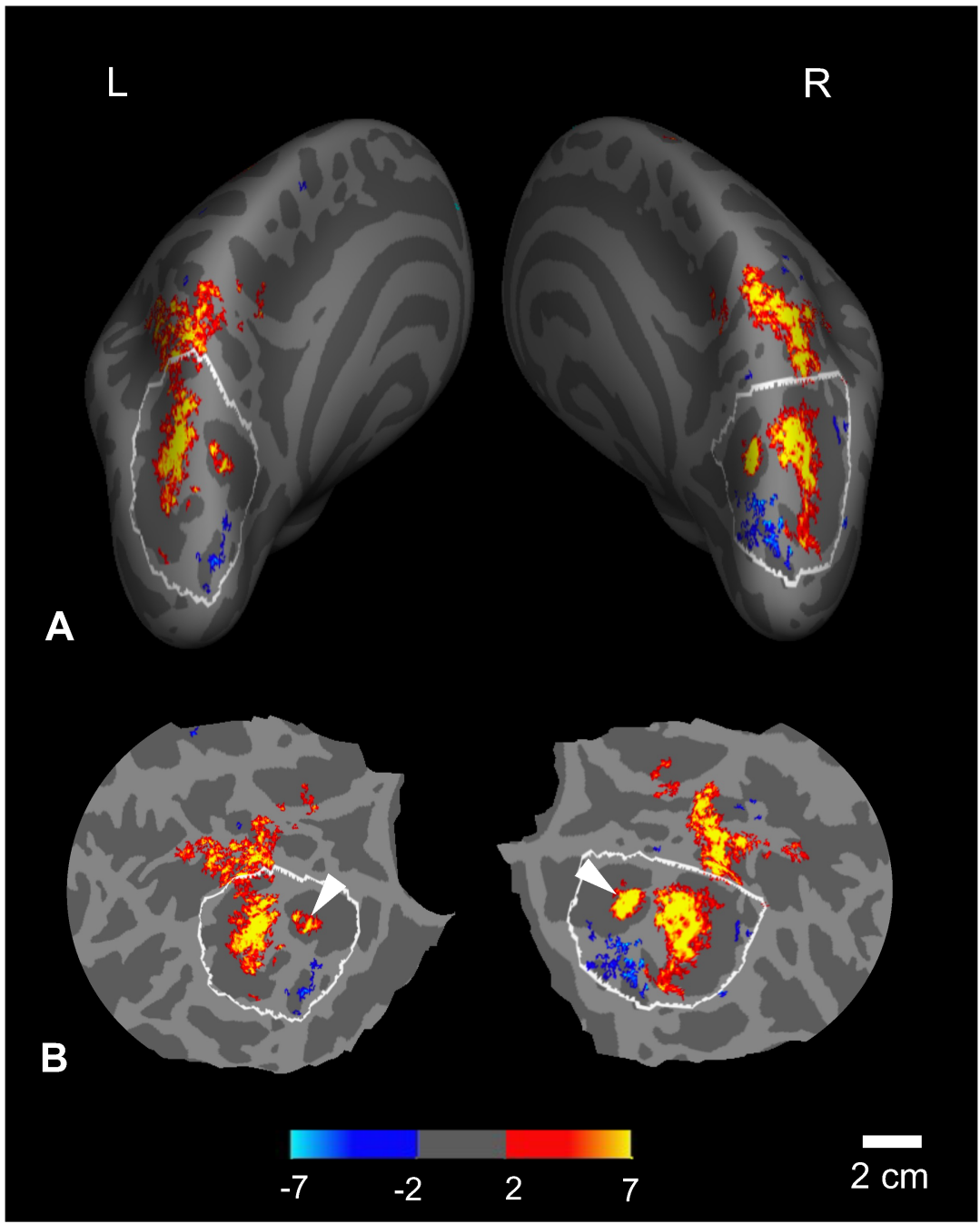

**Figure 2.**

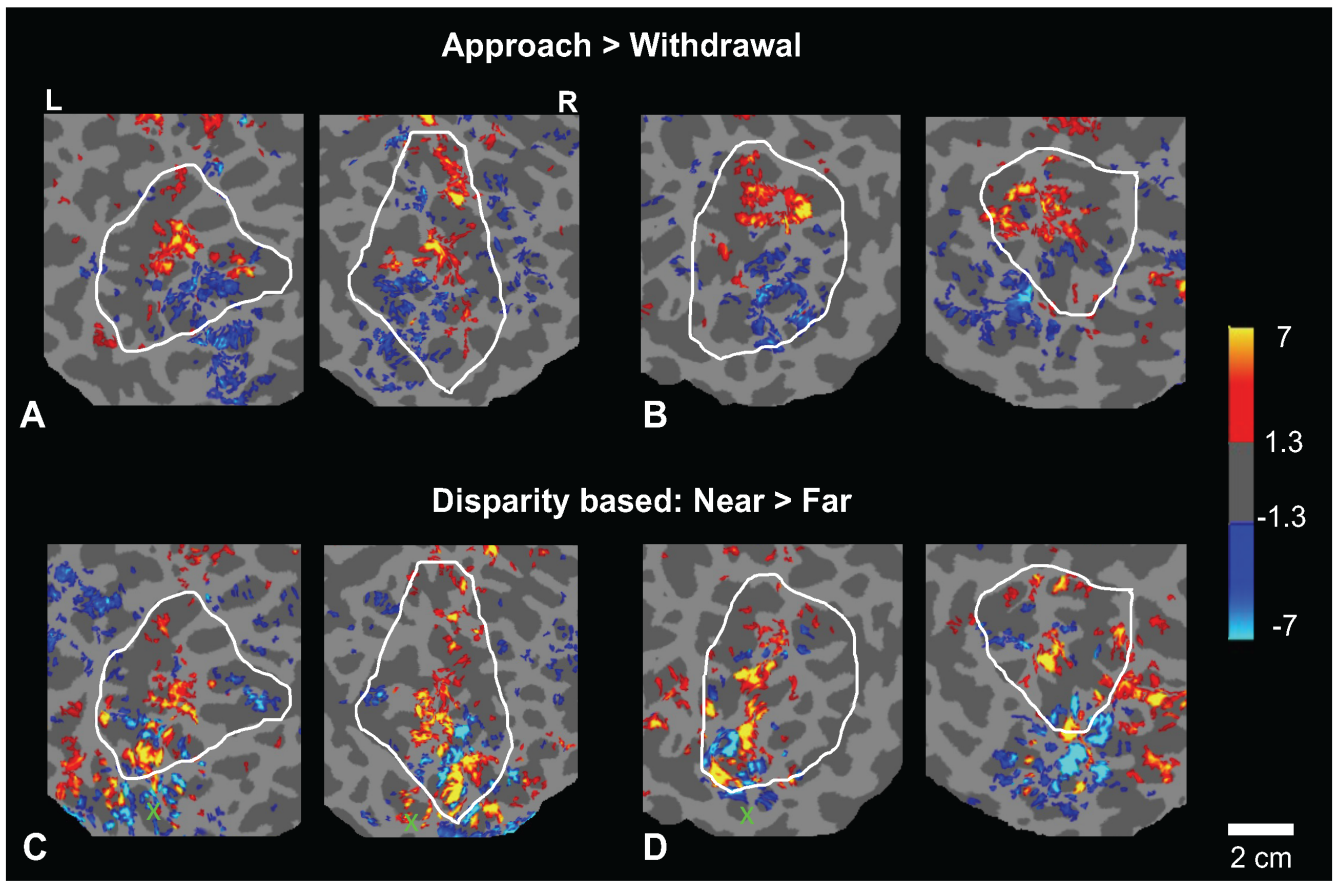

Figure 3.

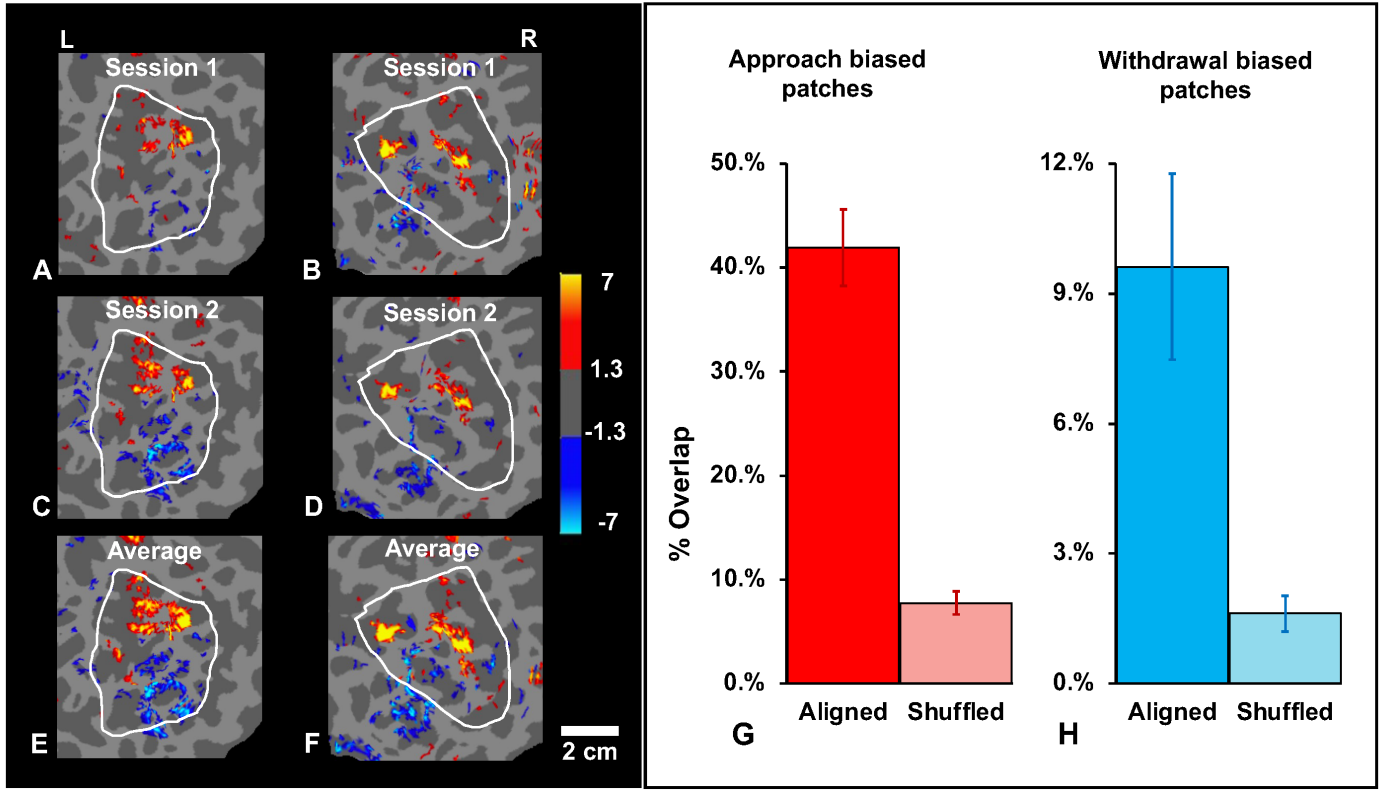

Figure 4.

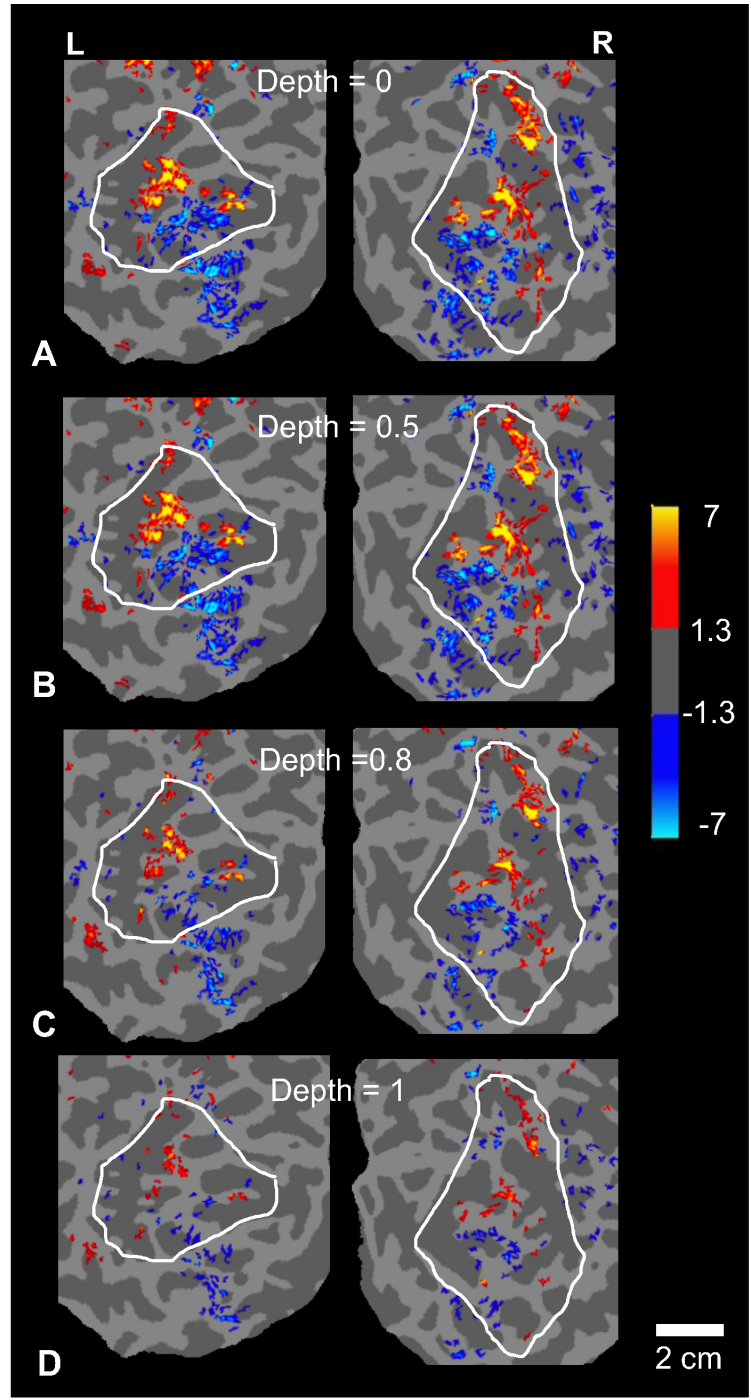

**Figure 5.**

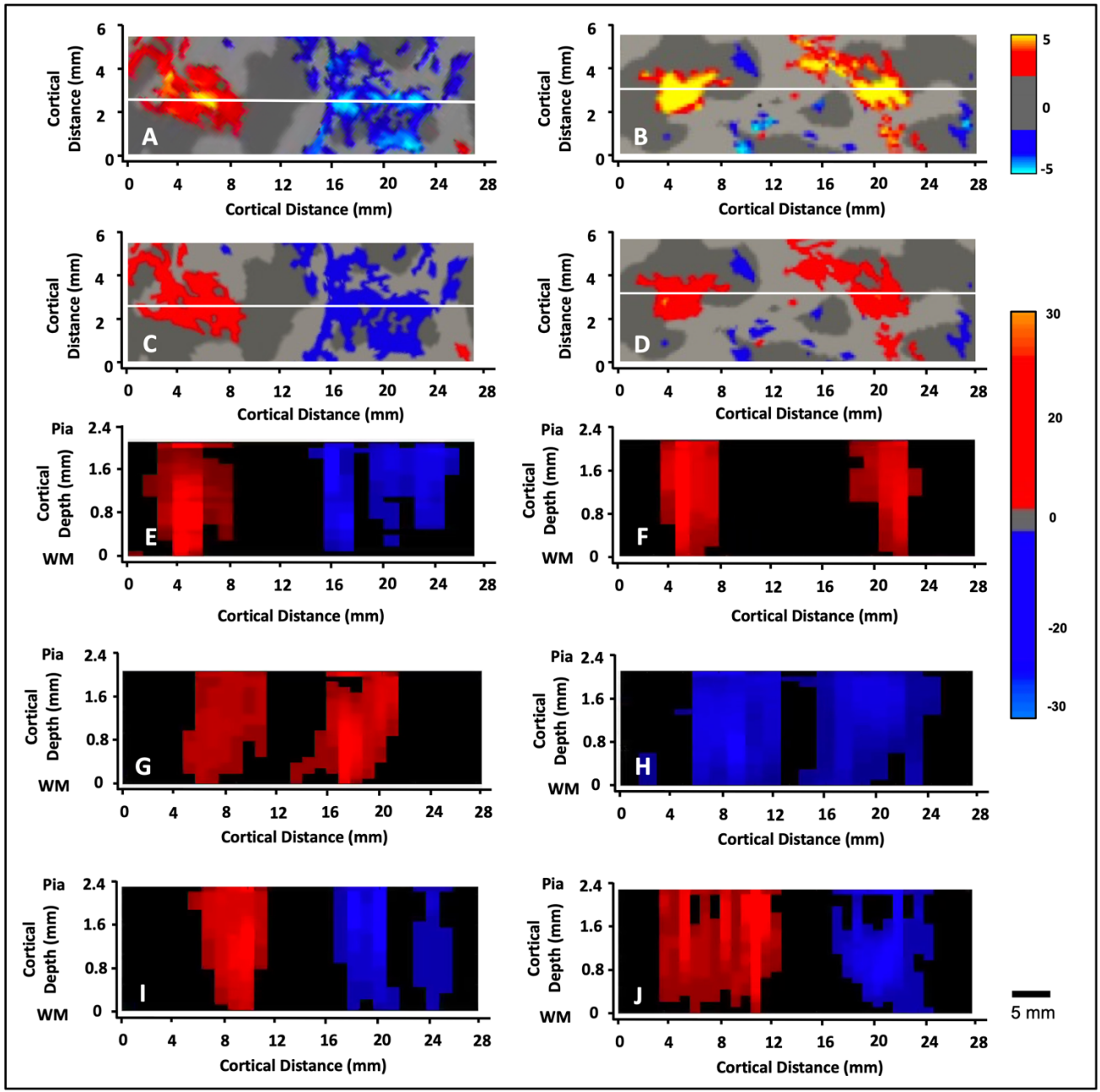

**Figure 6.**

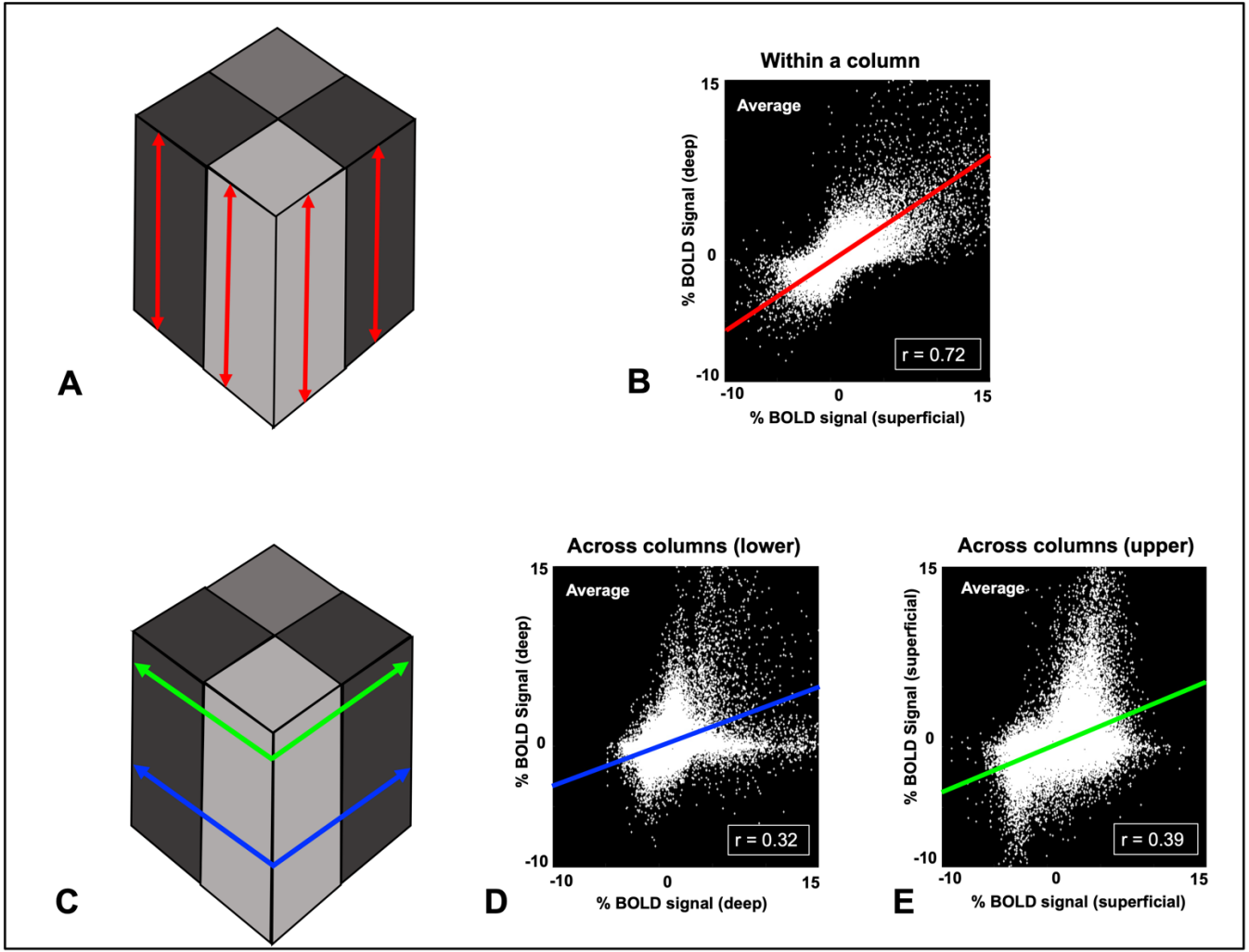

**Figure 7.**

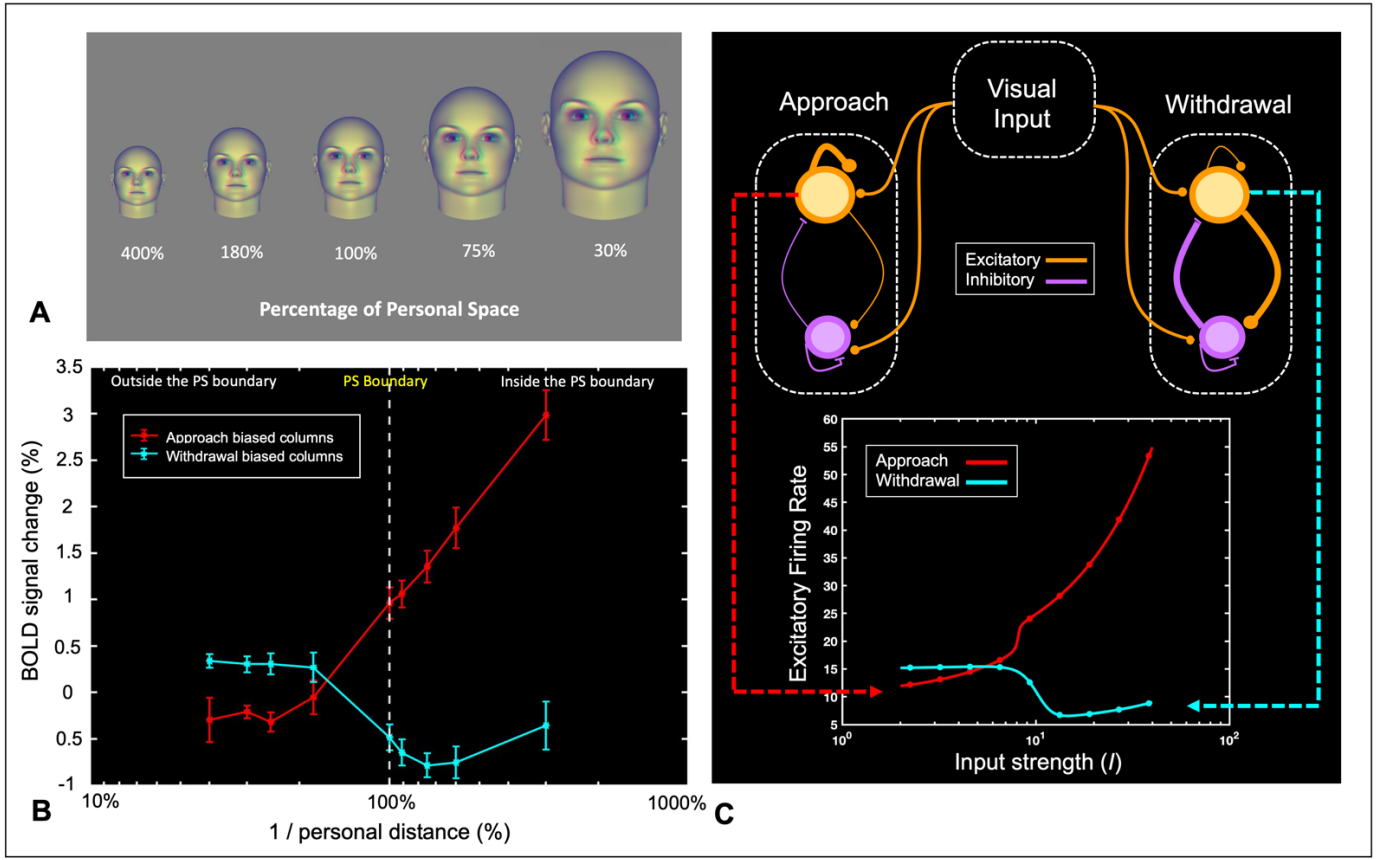

**Figure 8.**

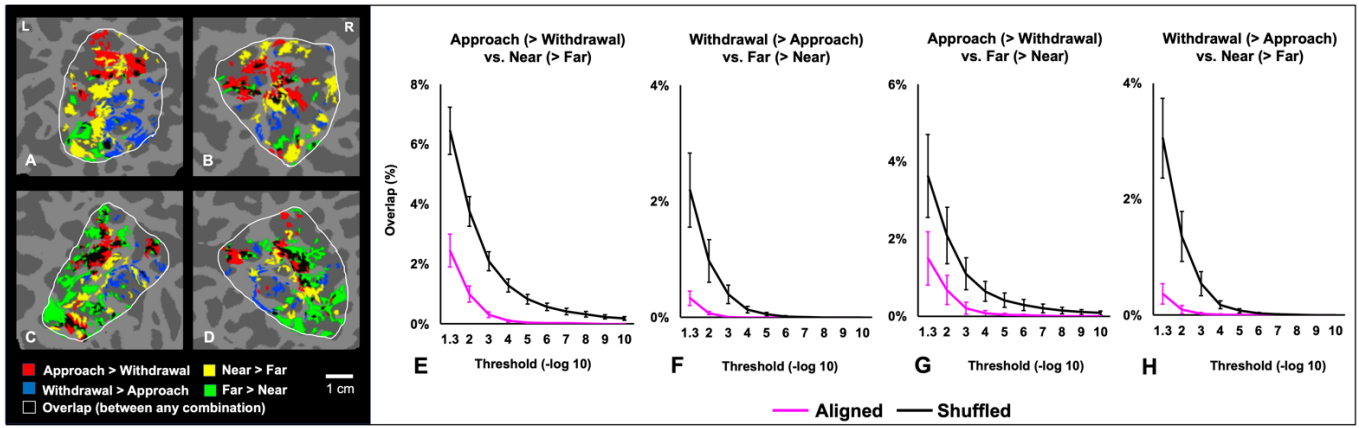

**Figure 9.**

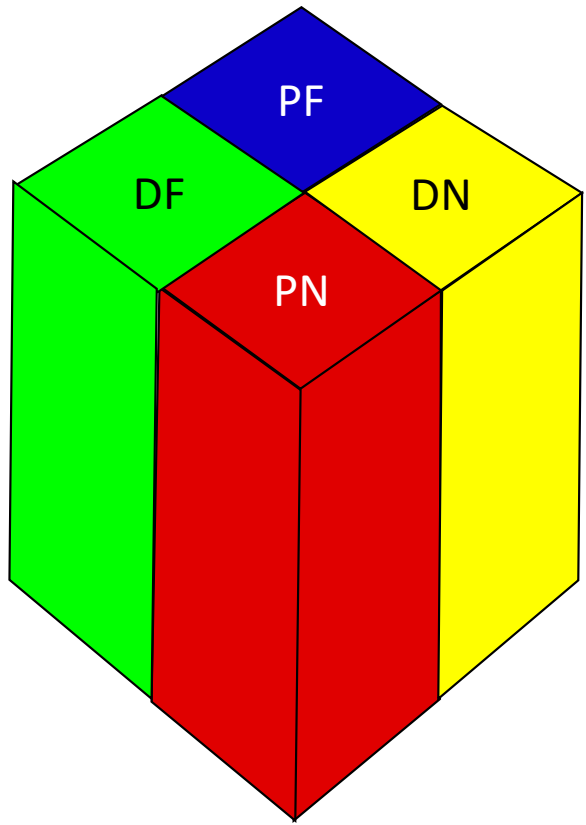

**Figure 10.**

| Parameter | Description | Approach | Withdrawal |
|---|---|---|---|
| $r_E$ | Excitatory (E) firing rate | | |
| $r_I$ | Inhibitory (I) firing rate | | |
| $W_{EE}$ | E-to-E synaptic weight | 5 | 0.1 |
| $W_{EI}$ | I-to-E synaptic weight | -0.1 | -10 |
| $W_{IE}$ | E-to-I synaptic weight | 0.1 | 10 |
| $W_{II}$ | I-to-I synaptic weight | -1 | -0.1 |
| $\theta_E$ | E baseline input | 10 | 10 |
| $\theta_I$ | I baseline input | 10 | 1 |
| $W_E$ | Input synaptic weight to E | 1 | 0.1 |
| $W_I$ | Input synaptic weight to I | 10 | 1 |
| $\theta$ | Neural threshold | 20 | |
| $I$ | Input | | |

**Table 1.**